\documentclass[preprint,%showpacs,
preprintnumbers,amsmath,amssymb,nofootinbib]{revtex4}

\usepackage{etex}
\usepackage{amssymb,amsthm,amscd,amsbsy,array}
\usepackage{bm}% bold math
\usepackage{soul} % underline, strikethrough, etc.
\usepackage{graphics,graphicx,xcolor}
\usepackage{soul} % underline, strikethrough,
\usepackage{algorithm}
\usepackage{algorithmic}
\usepackage{longtable}
\usepackage{graphicx}
\usepackage[caption=false]{subfig}

\usepackage{epstopdf}
\usepackage{float}
\usepackage[utf8]{inputenc}
\usepackage[T1]{fontenc}
\usepackage[english]{babel}
\usepackage{graphicx}
\usepackage{float}
\usepackage{tikz}
\usepackage{xcolor}
\usepackage[utf8]{inputenc}
\usepackage[T1]{fontenc}
\usepackage[english]{babel}
\usepackage{graphicx}
\usepackage{float}
\usepackage{tikz}
\usepackage{xcolor}
\usepackage{makecell}
\usepackage{esint}
\usepackage{cancel} 
\usepackage{blkarray}
\usepackage{mathrsfs}
\usepackage[normalem]{ulem}
\usepackage[colorlinks=true, pdfstartview=FitV, linkcolor=blue, citecolor=blue, urlcolor=blue]{hyperref} % hyperref

\newcommand{\red}{\textcolor{red}}

\newcommand{\blue}{\textcolor{blue}}
\newcommand{\green}{\textcolor{green}}

\newcommand{\purple}{\textcolor[rgb]{0.5,0.0,0.5}}
\newcommand{\gb}{\quad\colorbox{green}}

\newenvironment{redtext}{\color{red}}
{\ignorespacesafterend}
\newenvironment{bluetext}{\color{blue}}{\ignorespacesafterend}

\newenvironment{magentatext}{\color{magenta}}{\ignorespacesafterend}
\newenvironment{cyantext}{\color{cyan}}{\ignorespacesafterend}
\newenvironment{orangetext}{\color{orange}}
{\ignorespacesafterend}

\newcommand{\bmagenta}{\begin{magentatext}}
	\newcommand{\emagenta}{\end{magentatext}}
\newcommand{\bcyan}{\begin{cyantext}}
	\newcommand{\ecyan}{\end{cyantext}}
\newcommand{\bblue}{\begin{bluetext}}
	\newcommand{\eblue}{\end{bluetext}}
\newcommand{\bred}{\begin{redtext}}
	\newcommand{\ered}{\end{redtext}}
\newcommand{\borange}{\begin{orangetext}}
	\newcommand{\eorange}{\end{orangetext}}

\numberwithin{equation}{section}

\let\ssection=\section
\renewcommand{\section}{\setcounter{equation}{0}\ssection}
\newcommand{\beq}{\begin{equation}}
	\newcommand{\eeq}{\end{equation}}

\def\s.t.{{\quad\text{\small such that}\quad}}

\renewcommand{\Im}{\mathrm{Im}}

\renewcommand{\Re}{\mathrm{Re}}

\def\smallover\#1/\#2{\hbox{$\textstyle\frac{\#1}{\#2}$}} %

\def\benu{\begin{enumerate}}
	\def\eenu{\end{enumerate}}
\def\bitem{\begin{itemize}}
	\def\eitem{\end{itemize}}

\def\beq{\begin{equation}}
	\def\eeq{\end{equation}}
\def\beqa{\begin{eqnarray}}
	\def\eeqa{\end{eqnarray}}

\def\barray{\left(\begin{array}}
	\def\earray{\end{array}\right)}
\def\barraynb{\begin{array}}
	\def\earraynb{\end{array}}

\def\?{{\quad\gb{\fbox{\texttt{?}}\;}}\quad}

\def\v0{\mathbf{0}}

\def\smallover#1/#2{\hbox{$\textstyle\frac{#1}{#2}$}} %
\def\smallcirc{{\raise 0.5pt \hbox{$\scriptstyle\circ$}}}
\def\cabove(#1){\stackrel{\smallcirc}{#1}}
\def\ccabove(#1){\stackrel{\smallcirc\smallcirc}{#1}}
\def\cccabove(#1){\stackrel{\,\smallcirc\smallcirc\smallcirc}{#1}\,}
\def\2{{\smallover1/2}}

\def\ie{{\;\text{\small i.e.}\;}}
\def\ie,{{\;\text{\small i.e.,}\;}}

\let\ssection=\section
\renewcommand{\section}
{\setcounter{equation}{0}\ssection}

\makeatletter
\renewcommand{\fnum@figure}{FIG.~\thefigure}
\makeatother

\begin{document}

	\title{Decay criteria for asymptotic freedom in plane gravitational waves
	}
	
	\author{
		Qi-Liang Zhao$^{1}$\footnote{mailto: zhaoqliang@ucas.ac.cn}
		Li-Ming Cao$^{2}$\footnote{mailto: caolm@ustc.edu.cn}
	}
	
	\affiliation{
		${}^1$ School of Fundamental Physics and Mathematical Sciences,
		Hangzhou Institute for Advanced Study, UCAS, Hangzhou 310024, China
		\\
		${}^2$ Interdisciplinary Center for Theoretical Study and Department of Modern Physics,
		University of Science and Technology of China, Hefei 230026, China
		%%%%%%%
	}
	\date{\today}

	%}

	\begin{abstract}
		We investigate when plane-wave memory admits standard outgoing free data beyond the idealized sandwich-wave approximation. For a Brinkmann plane wave with profile $A(U)$, the commonly used condition $A(U)|_{U\to\infty}=0$ is not sufficient to guarantee ordinary asymptotically free motion. From the integral form of the transverse geodesic equation, we derive weighted decay criteria which divide the asymptotic dynamics into strongly asymptotically free, weakly asymptotically free, and non-asymptotically free motions. These motions are realized explicitly by the new analytical solutions of three typical examples: a Scarf profile, an inverse-cubic profile, and an inverse-square profile. A surprising feature is that the drift correction in the weakly asymptotically free motion affects only trajectories with nonzero outgoing velocity and therefore does not obstruct displacement memory. We further express the classification in terms of the accumulated tidal matrix, showing that it is an intrinsic curvature effect rather than a coordinate artifact.
	\end{abstract}
	
	\maketitle
	\tableofcontents

	%%%%%%%%%%%%%%%%%%%%%%%%%%
	
	\section{Introduction}
	
	Gravitational-wave memory is one of the key predictions arising from the nonlinear nature of general relativity \cite{Christodoulou:1991cr}. It refers to a persistent change in the relative configuration of freely falling test particles after the passage of gravitational radiation \cite{Zeldovich:1974gvh}. In asymptotically flat spacetimes, gravitational memory is deeply tied to the structure of future null infinity $\mathscr{I}^+$ \cite{Strominger:2014pwa,Mitman:2024uss}. The nonlinear displacement memory can be understood as a transition between distinct Bondi vacua, or equivalently as a change in the asymptotic shear \cite{Mitman:2024uss}. This interpretation is closely related to BMS supertranslations, soft graviton theorems, and flux-balance laws \cite{He:2014laa,Strominger:2014pwa,Mitman:2024uss}. From a more geometric viewpoint, $\mathscr{I}^+$ carries a conformal Carrollian structure \cite{Duval:2014uva}, and recent developments have shown that the Bondi flux-balance equations can be derived from local Carrollian, Weyl, and diffeomorphism invariance at the null conformal boundary \cite{Fiorucci:2025twa}. These results provide a boundary-geometric explanation of how radiative degrees of freedom source the evolution of Bondi charges. At the same time, they also make clear that memory observables at $\mathscr{I}^+$ are sensitive to the choice of BMS frame \cite{Mitman:2024uss,Cristofoli:2025esy}.
	
	Memory effects, however, are not only restricted to the conventional asymptotically flat formulation. Plane gravitational waves also provide an exact and highly symmetric arena in which memory can be studied nonperturbatively \cite{Zhang:2017rno,Zhang:2017geq}. In Brinkmann coordinates, a generally polarized plane gravitational wave (GW) can be written as
	\begin{eqnarray}
		ds^2=dX^2+dY^2+2dUdV-kA(U)(X^2-Y^2)dU^2,
		\label{metric-PGW}
	\end{eqnarray}
	where $X$ and $Y$ are coordinates on the wave-front plane; $U=\frac{T-Z}{\sqrt{2}}$ and $V=\frac{T+Z}{\sqrt{2}}$ are light-cone coordinates; $A(U)$ is the profile for the ``$+$'' polarization modes with constant amplitude $k$. Here we ignore the other ``$\times$'' polarization since the memory effect is primarily concerned with the detection of non-precessing binaries, and in such systems the contribution of the memory is mainly in the ``$+$'' polarization mode \cite{Rossello-Sastre:2024zlr,Favata:2008yd}.
	
\begin{figure}[htbp]
		\centering

		\tikzset{every picture/.style={line width=0.75pt,scale=.9}} %set default line width to 0.75pt        
		
		\begin{tikzpicture}[x=0.75pt,y=0.75pt,yscale=-1,xscale=1]
			
			%Shape: Rectangle [id:dp6220526049291921] 
			\draw  [color={rgb, 255:red, 255; green, 255; blue, 255 }  ,draw opacity=1 ][fill={rgb, 255:red, 245; green, 166; blue, 35 }  ,fill opacity=1 ] (176.28,211.9) -- (422.18,30.14) -- (467.92,92.03) -- (222.02,273.79) -- cycle ;
			%Shape: Rectangle [id:dp45621496278657725] 
			\draw  [color={rgb, 255:red, 144; green, 19; blue, 254 }  ,draw opacity=1 ][fill={rgb, 255:red, 144; green, 19; blue, 254 }  ,fill opacity=1 ] (176.28,211.9) -- (422.18,30.14) -- (426.34,35.77) -- (180.43,217.53) -- cycle ;
			%Shape: Rectangle [id:dp7371227960228692] 
			\draw  [color={rgb, 255:red, 144; green, 19; blue, 254 }  ,draw opacity=1 ][fill={rgb, 255:red, 144; green, 19; blue, 254 }  ,fill opacity=1 ] (218.94,269.63) -- (464.85,87.87) -- (467.92,92.03) -- (222.02,273.79) -- cycle ;
			%Straight Lines [id:da9344375897630053] 
			\draw [line width=2.25]  [dash pattern={on 6.75pt off 4.5pt}]  (200.1,241.47) -- (444.1,62.47) ;
			%Straight Lines [id:da4959902801506828] 
			\draw [color={rgb, 255:red, 74; green, 144; blue, 226 }  ,draw opacity=1 ][line width=1.5]    (402.47,165.14) -- (468.19,248.77) ;
			\draw [shift={(470.67,251.92)}, rotate = 231.84] [fill={rgb, 255:red, 74; green, 144; blue, 226 }  ,fill opacity=1 ][line width=0.08]  [draw opacity=0] (11.61,-5.58) -- (0,0) -- (11.61,5.58) -- cycle    ;
			\draw [shift={(400,162)}, rotate = 51.84] [fill={rgb, 255:red, 74; green, 144; blue, 226 }  ,fill opacity=1 ][line width=0.08]  [draw opacity=0] (11.61,-5.58) -- (0,0) -- (11.61,5.58) -- cycle    ;
			%Straight Lines [id:da1551924227102376] 
			\draw [color={rgb, 255:red, 74; green, 144; blue, 226 }  ,draw opacity=1 ][line width=1.5]    (194.47,49.14) -- (260.19,132.77) ;
			\draw [shift={(262.67,135.92)}, rotate = 231.84] [fill={rgb, 255:red, 74; green, 144; blue, 226 }  ,fill opacity=1 ][line width=0.08]  [draw opacity=0] (11.61,-5.58) -- (0,0) -- (11.61,5.58) -- cycle    ;
			\draw [shift={(192,46)}, rotate = 51.84] [fill={rgb, 255:red, 74; green, 144; blue, 226 }  ,fill opacity=1 ][line width=0.08]  [draw opacity=0] (11.61,-5.58) -- (0,0) -- (11.61,5.58) -- cycle    ;
			%Shape: Axis 2D [id:dp16981488656253452] 
			\draw  (327.67,276.59) -- (410,276.59)(335.9,200.92) -- (335.9,285) (403,271.59) -- (410,276.59) -- (403,281.59) (330.9,207.92) -- (335.9,200.92) -- (340.9,207.92)  ;
			%Straight Lines [id:da8952533956880002] 
			\draw    (506,135) -- (596.67,134.92) ;
			\draw [shift={(599.67,134.92)}, rotate = 179.95] [fill={rgb, 255:red, 0; green, 0; blue, 0 }  ][line width=0.08]  [draw opacity=0] (8.93,-4.29) -- (0,0) -- (8.93,4.29) -- cycle    ;
			%Straight Lines [id:da5514528337739152] 
			\draw    (115,131) -- (205.67,130.92) ;
			\draw [shift={(112,131)}, rotate = 359.95] [fill={rgb, 255:red, 0; green, 0; blue, 0 }  ][line width=0.08]  [draw opacity=0] (8.93,-4.29) -- (0,0) -- (8.93,4.29) -- cycle    ;
			
			% Text Node
			\draw (267.04,161.91) node [anchor=north west][inner sep=0.75pt]  [rotate=-322.3] [align=left] {Wave zone};
			% Text Node
			\draw (449,45.4) node [anchor=north west][inner sep=0.75pt]    {$U=0$};
			% Text Node
			\draw (425,10.4) node [anchor=north west][inner sep=0.75pt]    {$U=U_{A}$};
			% Text Node
			\draw (472,85.4) node [anchor=north west][inner sep=0.75pt]    {$U=U_{B}$};
			% Text Node
			\draw (345,200.4) node [anchor=north west][inner sep=0.75pt]    {$T$};
			% Text Node
			\draw (417,269.4) node [anchor=north west][inner sep=0.75pt]    {$Z$};
			% Text Node
			\draw (234,50) node [anchor=north west][inner sep=0.75pt]   [align=left] {Before-zone};
			% Text Node
			\draw (455,178) node [anchor=north west][inner sep=0.75pt]   [align=left] {After-zone};
			% Text Node
			\draw (127,138) node [anchor=north west][inner sep=0.75pt]   [align=left] {Downwind};
			% Text Node
			\draw (528,143) node [anchor=north west][inner sep=0.75pt]   [align=left] {Upwind};

		\end{tikzpicture}
		
		\caption{
			\textit{\small 
				This figure shows the profile of a sandwich wave. It consists of three parts: the before- and after-zones correspond, respectively, to the flat spacetime before the arrival and after the passage of the gravitational wave, where the test particles undergo free motion; the wave-zone describes the region of gravitational wave interaction.
			}
			\label{sandwich-plot}}
		
	\end{figure}
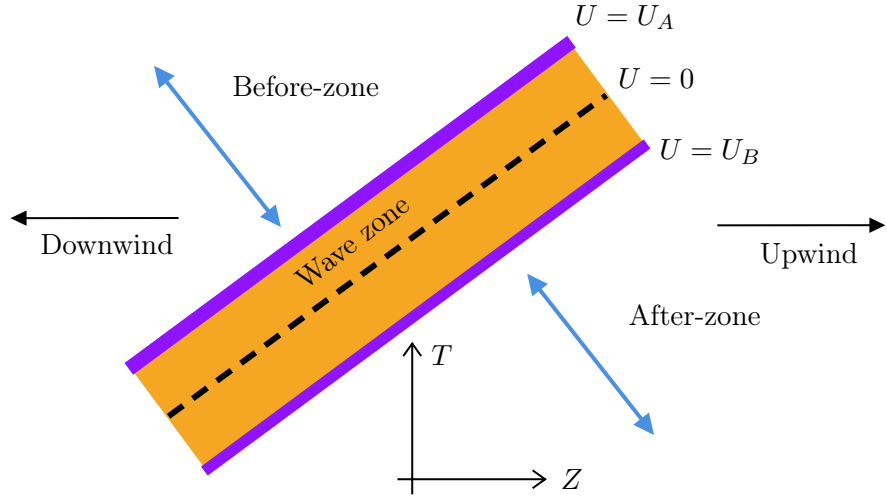

	The memory effect for plane gravitational waves was systematically studied by Zhang, Duval, Gibbons and Horvathy in the sandwich-wave setting \cite{Zhang:2017rno}. For a compactly supported sandwich profile, the spacetime naturally decomposes into an incoming before-, wave-, and outgoing after-zones, as shown in FIG. \ref{sandwich-plot}.
	
	A test particle initially at rest before the wave generally emerges with a nonzero constant velocity after the wave has passed; this is the velocity memory effect \cite{Zhang:2017rno,Zhang:2018srn,Carneiro:2026iuk}. In special cases, the outgoing  velocity may vanish while a finite displacement remains \cite{Zhang:2024uyp}. This displacement memory arises as a special case of the more general velocity memory phenomenon \cite{Zhang:2024uyp,Zhang:2024tey,Zhang:2025lxs}. It should be distinguished conceptually from the vacuum-transition displacement memory familiar at $\mathscr{I}^+$, although possible relations between these notions remain an interesting question.
	
	Sandwich plane waves and their associated velocity memory have recently attracted renewed attention as exact backgrounds for scattering-amplitude and waveform calculations \cite{Adamo:2022qci,Cristofoli:2025esy}. In these works, the plane wave is treated as a nonlinear gravitational background, and the memory data of this background enter the gravitational waveform emitted by a massive particle moving on it \cite{Adamo:2022qci}. In particular, the velocity-memory matrix appears in the boundary conditions of the external states and affects the waveform through the transverse vielbein data, Synge's world function, and Green-function tail terms \cite{Cristofoli:2025esy}. This shows that plane-wave velocity memory is not merely a feature of geodesic motion in the bulk, but can also become part of radiative observables. At the same time, these analyses rely on the existence of well-defined incoming and outgoing scattering data, which is usually guaranteed by imposing a sandwich profile or by assuming sufficiently controlled asymptotic regions. For non-compact plane-wave profiles, however, the precise decay conditions required for such data to exist remain to be clarified.
	
	A more direct motivation for our work comes from recent approximate studies of plane-wave memory \cite{Zhao:2026zpd}. It was observed in Ref. \cite{Zhao:2026zpd} that different sandwich or rapidly decaying profiles can lead to similar outgoing particle trajectories, even when their detailed structures inside the wave-zone are rather different. This observation suggests that the late-time motion is not controlled primarily by the detailed shape of the profile in the wave-zone, but by the effect of its asymptotic tail. However, this conclusion was obtained for profiles whose decay is already fast enough to ensure ordinary outgoing free motion, such as compactly supported or exponentially decaying profiles. It therefore does not identify the critical decay behavior separating well-defined free out-data from long-range or non-free asymptotics.
	
	Motivated by these recent studies, we ask what remains true beyond the sandwich approximation. In explicit calculations one often replaces the compact sandwich profile by a smooth non-compact function satisfying
	\begin{eqnarray}
		A(U)\bigg|_{U\to\pm\infty}=0.
		\label{flat-cond}
	\end{eqnarray}
	However, this weaker condition does not by itself guarantee standard free asymptotics for test particles. A slowly decaying tail can accumulate and produce modified asymptotic behavior. Therefore, one must impose stronger conditions on the profile in order to ensure the existence of outgoing free data. This is the central question addressed in this paper.
	
	The purpose of this paper is therefore to clarify the prior dynamical question on which such scattering descriptions often rely: when are the free in/out data and the corresponding velocity-memory data well defined for a non-sandwich plane GW profile? In this sense, our results extend the geometric understanding of plane-wave memory beyond the compactly supported case. They also provide a profile-level classification of short-range and long-range plane-wave tails in terms of the asymptotic behavior of test-particle motion.
	
	The paper is organized as follows. In Sec. \ref{Sec-cond}, we derive the general outgoing behavior of test particles induced by the accumulated plane-wave profile, formulate refined decay conditions on $A(U)$, and classify the outgoing motion into strongly asymptotically free, weakly asymptotically free, and non-asymptotically free motions. In Sec. \ref{three-examples}, we illustrate these three kinds of asymptotic motions using explicit solvable examples. In Sec. \ref{Intrinsic Physical Perspective}, we discuss the intrinsic meaning of the classification in terms of the tidal matrix and the geodesic-deviation equation, emphasizing that the results are not artifacts of a particular coordinate choice. Sec. \ref{Summary} contains our summary.
	
	\section{Motion of a test particle and the Restriction on the wave profile for free motion}\label{Sec-cond}
	
	In this section, we first present the general integral solution of the geodesic equation for plane GWs. Based on this integral solution, we then demonstrate that the type of asymptotic motion of a test particle after the passage of a sandwich GW depends solely on the tail of the profile, and is independent of the oscillations in the wave-zone, the latter only determines the initial velocity and position of the outgoing data. Finally, we provide a classification of the asymptotic motion based on the cumulative effect of the profile tail.
	
	\subsection{General solution of the geodesic equation in integral form}
	
	When $U$ is taken as the affine parameter, the transverse geodesics in the plane GW background \eqref{metric-PGW} can be reduced to the motion of an anisotropic 2-dimensional oscillator, while the longitudinal coordinate $V$ can be expressed in terms of the action of the transverse motion \cite{Zhang:2017rno}. In this way the  motion of a single particle is described by the following equations
	\begin{eqnarray}
		&&
		\frac{d^2}{dU^2}X^i(U)+(-1)^{i+1} \ k A(U)X^i(U)=0, 
		\label{geo-X-i} \\
		&&
		\frac{d}{dU}V(U)=-L+\varepsilon, \quad L=\frac{1}{2}\left(\frac{d\bm{X}}{dU}\right)^2-\frac{1}{2}kA(U)(X^2-Y^2),
		\label{geo-V}
	\end{eqnarray}
	where $\varepsilon \in \mathbb{R}$ and $\varepsilon > 0$, $< 0$, $= 0$ give space-like, time-like, and null geodesics, respectively; $i=1,2$ indicate the coordinates of transverse plane $X$ and $Y$, respectively. 

	Since the independent dynamical degrees of freedom are the transverse components, we mainly restrict our attention to these directions.
	
	Integrating \eqref{geo-X-i} directly we have the solution for the particle motion initially at rest
	\begin{eqnarray}
		X^{i}=X^i_0+(-1)^i \ k\int_{U_0}^{U}ds \ (U-s)A(s)X^i(s),
		\label{geo-sol-i}
	\end{eqnarray}
	where $U_0$ is the initial time, $X^i_0$ denotes the initial position of the test particle. The calculation details can be found in Appendix. \ref{Cal-Int-sol}.
	
	\subsection{Asymptotic motion types: Irrelevance of the wave-zone and sole dependence on the profile tail}
	
	It is worth mentioning that the integral formulation in \eqref{geo-sol-i} also makes clear why the late-time classification of the particle motion is controlled by the asymptotic tail of the profile rather than by the detailed structure of the wave zone, as mentioned in the summary of Ref. \cite{Zhao:2026zpd}. 
	
	For definiteness, let $U_*$ be a large but finite value beyond which the profile has entered its asymptotic decay regime. Then we can split the integral solution of \eqref{geo-sol-i} into two intervals. The first interval $[U_0,U_*]$ represents the segment from the initial time, through the gravitational wave interaction, up to the point where the tail decay begins; it includes the wave-zone region of the sandwich profile. The other interval $[U_*,\infty]$ represents the region after the tail decay begins at $U_*$. Under this splitting, \eqref{geo-sol-i} can be rewritten as
	\begin{eqnarray}
		&&
		X^i(U)=X^i_*+V^i_*(U-U_*)
		+(-1)^{i}k\int_{U_*}^{U}ds\,(U-s)A(s)X^i(s), 
		\label{no-relation-wavezone-1}
	\end{eqnarray}
	where $U>U_*$, $X^i_*$ is the initial position upon entering the asymptotic decay region, resulting from the cumulative effect over the interval from the initial time $U_0$ to the time $U_*$ when the tail decay begins,
	\begin{eqnarray}
		X^i_*=X^i_0+(-1)^{i}k\int_{U_0}^{U_*}ds \ (U_*-s) \ A(s)X^i(s);
	\end{eqnarray}
	and $V^i_*$ is the initial velocity in the asymptotic decay region, which likewise originates from the cumulative effect over the interval $[U_0,U_*]$,
	\begin{eqnarray}
		V^i_*=(-1)^ik\int_{U_0}^{U_*}ds\left[A(s)X^i(s)\right].
	\end{eqnarray}
	
	Here we can see that, all contributions from the finite interval $[U_0,U_*]$, including the detailed oscillatory behavior of the wave zone, are absorbed into the effective data $(X^i_*,V^i_*)$. Provided the wave-zone contribution is finite, as is automatic for smooth profiles on finite intervals and remains true for piecewise continuous or locally integrable profiles, it cannot change the tail-controlled asymptotic universality class of the profile. It only changes the numerical values of the outgoing data. The distinction between different late-time regimes is therefore determined by the tail integral over $U>U_*$.
    
    \subsection{Classification of asymptotic motion based on the decay rate of $A(U)$}
	
	Note that \eqref{no-relation-wavezone-1} can be rewritten as
	\begin{eqnarray}
		X^{i}=X^i_*+(V^i_*+\Delta V^i)U+\Delta X^i,
		\label{int-sol-1}
	\end{eqnarray}
	where $\Delta V^i$ is the accumulated correction of the final velocity, and $\Delta X^i$ denotes the accumulated correction to the free-line intercept, respectively,
	\begin{eqnarray}
		&&
		\Delta V^i=(-1)^i \ k\int_{U_*}^{U}ds \ \left[A(s)X^i(s)\right], 
		\\
		&&
		\Delta X^i=(-1)^{i+1} \ k\int_{U_*}^{U}ds \ \left[sA(s)X^i(s)\right]-V^i_*U_*.
		\label{int-sol-2}
	\end{eqnarray}
	
	If the corresponding limits exist as $U\to\infty$,
	\begin{eqnarray}
		\int_{U_*}^{\infty}ds \ \left[A(s)X^i(s)\right]<\infty, \quad
		\int_{U_*}^{\infty}ds \ \left[sA(s)X^i(s)\right]<\infty.
		\label{int-A-X}
	\end{eqnarray}
	then the motion admits ordinary free out-data.
	
	However, the integrals in \eqref{int-A-X} all involve $X^i(s)$, whose explicit form requires solving the geodesic equation \eqref{geo-X-i}. To eventually arrive at a convergence condition for the integrals that depends solely on $A(U)$, we need to find a way to simplify, or even eliminate its influence.
	
	Since the integration interval in \eqref{int-A-X} lies in the tail region, if ordinary outgoing free data exist, the leading asymptotic form must be
    \begin{eqnarray}
    	X^i(U)\bigg|_{U\to\infty}\sim a^i+b^iU+o(1).
    	\label{free-X-i}
    \end{eqnarray}
    Substituting this trial free behavior into \eqref{int-A-X} we identify the weighted tail integrals that control the existence of the velocity and intercept
    \begin{eqnarray}
    	\int_{U_*}^{\infty}ds \ \left(s\left|A(s)\right|\right)<\infty, \quad
    	\int_{U_*}^{\infty}ds \ \left(s^2\left|A(s)\right|\right)<\infty.
    	\label{A-cond}
    \end{eqnarray}
    Here we impose absolute convergence of the weighted tail integrals for the following reasons: for oscillatory tails, conditional cancellations may improve the asymptotic behavior. For example, $A(U)\sim\frac{\sin U}{U^2}$ does not satisfy the absolute first-moment condition $\int_{U_*}^{\infty}s|A(s)|ds<\infty$, although its oscillatory tail can still lead to finite free outgoing data. Such oscillatory profiles require a separate analysis and are not included in the non-oscillatory decay hierarchy considered in this paper.
    
    The two integrals in \eqref{A-cond} control two questions: whether the outgoing velocity can settle to a finite limit, and whether the trajectory admits a finite free-line intercept, respectively. Based on these two integrals of different strengths, we can classify the asymptotic motion of the particle into the following three types: 
    \begin{itemize}
    	\item 
    	\textbf{Strongly asymptotically free motion:}
    	
    	If $A(U)$ satisfies
    	\begin{eqnarray}
    		\int_{U_*}^{\infty}ds \ \left(s^2\left|A(s)\right|\right)<\infty,
    		\label{A-cond-1}
    	\end{eqnarray}
    	the asymptotic motion of a test particle is strongly asymptotically free, coinciding with the strictly asymptotically free solution in \eqref{free-X-i}; both the outgoing velocity and free-line intercept are well defined.
    	
    	\item 
    	\textbf{Weakly asymptotically free motion:}
    	
    	If $A(U)$ satisfies
    	\begin{eqnarray}
    		\int_{U_*}^{\infty}ds \ \left(s\left|A(s)\right|\right)<\infty, 
    		\quad \mathrm{but} \quad 
    		\int_{U_*}^{\infty}ds \ \left(s^2\left|A(s)\right|\right)\to\infty,
    		\label{A-cond-2}
    	\end{eqnarray}
    	the asymptotic motion is weakly asymptotically free. In this case, the asymptotic motion deviates from the baseline of free motion, implying that a drift arises.
    	
    	\item 
    	\textbf{Non-asymptotically free Motion}
    	
    	If $A(U)$ satisfies
    	\begin{eqnarray}
    		\int_{U_*}^{\infty}ds \ \left(s\left|A(s)\right|\right)\to\infty, 
    		\label{A-cond-3}
    	\end{eqnarray}
    	the test particle does not settle into free asymptotic motion; instead, its asymptotic behavior is governed by the specific fall-off of $A(U)$ and is strongly dependent on the value of the amplitude $k$.
    	
    \end{itemize} 
    
    In this precise sense, the detailed wave-zone evolution affects the scattering data but not the asymptotic universality class, which is fixed by the decay of the tail of $A(U)$.
    
    The three categories discussed above are also summarized in Table \ref{Table-1}, along with several representative examples.
    \begin{table}[htbp]
    	\centering
    	\caption{
    		Three categories of $A(U)$
    		\label{Table-1}
    	}
    	\begin{tabular}{|c|c|c|c|}
    		\hline
    		&   Integral condition & Examples & Critical decay rate
    		\\
    		\hline
    		\makecell{Strongly asymptotically \\ free motion} &
    		\makecell{$\int_{U_*}^{\infty}ds \ \left(s^2\left|A(s)\right|\right)<\infty$} & 
    		\makecell{$A(U)|_{U\gg1}\sim U^{-p}$, $p>3$; \\
    			or $A(U)|_{U\gg1}\sim e^{-U}$; \\
    			or faster decay} & 
    		---
    		\\
    		\hline
    		\makecell{Weakly asymptotically \\ free motion} & 
    		\makecell{$\int_{U_*}^{\infty}ds \ \left(s\left|A(s)\right|\right)<\infty$, \\
    			but  \\
    			$\int_{U_*}^{\infty}ds \ \left(s^2\left|A(s)\right|\right)\to\infty$} & 
    		\makecell{$A(U)|_{U\gg1}\sim U^{-p}$, $2<p\leq 3$} & 
    		\makecell{$A(U)|_{U\gg1}\sim U^{-3}$}
    		\\
    		\hline
    		\makecell{Non-asymptotically \\ free Motion} & 
    		\makecell{$\int_{U_*}^{\infty}ds \ \left(s\left|A(s)\right|\right)\to\infty$} & 
    		\makecell{$A(U)|_{U\gg1}\sim U^{-p}$, $0<p\leq2$} & 
    		\makecell{$A(U)|_{U\gg1}\sim U^{-2}$}
    		\\
    		\hline
    	\end{tabular}
    	
    \end{table}

	 In the next section, we will use three distinct exactly solvable examples to illustrate the detailed asymptotic motion corresponding to each of the three categories of $A(U)$.
	
	\section{Analytic Examples of the Three Different Types of Asymptotic Motion}\label{three-examples}
	
	We now consider three analytically solvable examples to illustrate the three different kinds of the asymptotic particle motion.
	
	\subsection{Strongly asymptotically free motion}\label{Strong-free}
	
\begin{figure}[htbp]
		\centering
		\includegraphics[scale=.5]{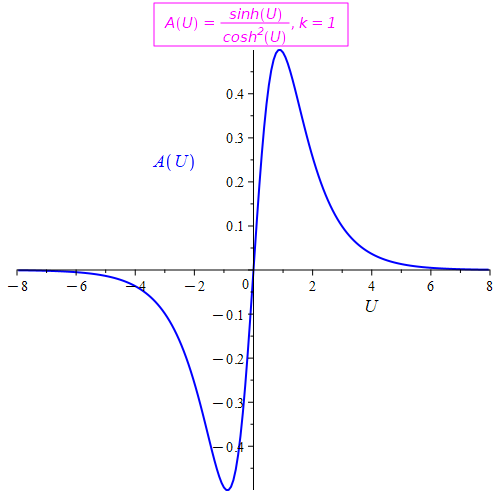}
		\caption{
			\textit{\small 
				This plot shows the shape of Scarf profile, and it rapidly falls off as $e^{-U}$ when $U\gg1$.
			}
		}
		\label{A-Scarf}
	\end{figure}

	The Scarf profile \cite{Zhang:2025lxs}, 
	\begin{eqnarray}
		A(U)=\frac{\sinh U}{\cosh^2U}, \label{Scarf}
	\end{eqnarray} 
	is a good example for strongly asymptotically free motion since $A(U)|_{U\gg1}\sim e^{-U}$. FIG. \ref{A-Scarf} shows the shape of this profile.
	
	We choose the Scarf profile as our first example for the following reasons. Although previous work has demonstrated the asymptotically free nature of the Scarf profile \cite{Zhang:2025lxs}, the discussion of its solutions, especially the two linearly independent solutions, remains incomplete. In this work, we obtain these two linearly independent solutions analytically by means of hypergeometric functions. They not only complete the earlier study, but are also of considerable value for studying the isometric Carroll symmetry involved \cite{Elbistan:2025wbp}. Moreover, as we will show, the hypergeometric method is in the same spirit as the calculations of most asymptotic behaviors in this paper. These considerations motivate our choice of the Scarf profile as the first case of study.
	
	\subsubsection{Analytical solution of the motion of a test particle}
	
	For the Scarf profile, the transverse geodesic equation can be written as 
	\begin{eqnarray}
		\frac{d^2}{dU^2}X^i(U)+(-1)^{i+1}k\frac{\sinh U}{\cosh^2U}X^i(U)=0.
		\label{geo-scarf}
	\end{eqnarray}
	By using 
	\begin{eqnarray}
		t=\sinh U, \label{t-U}
	\end{eqnarray}
	the above equation becomes \cite{Zhang:2025lxs}
	\begin{eqnarray}
		(1+t^2)\frac{d^2X^{i}}{dt^2}+t\frac{dX^{i}}{dt}+(-1)^{i+1} k\frac{t}{1+t^2}X^{i}=0.
		\label{eq-S-t}
	\end{eqnarray}
	
	\eqref{eq-S-t} can be solved in terms of the following two independent complex solutions
	\begin{eqnarray}
		X^i(t)=
		c_1 \ X^{i}_{+}(t)
		+
		c_2 \ X^{i}_{-}(t),
		\label{2sol-X-t-1}
	\end{eqnarray}
	where 
	\begin{eqnarray}
		X^{i}_{\pm}(t)=(t+I)^{\frac{1\mp\alpha^i}{4}} (t-I)^{\frac{1+\bar{\alpha}^i}{4}} \ F^i_{\pm}(t). \label{two-Hg-t} 
	\end{eqnarray}
	The function $F^i_{\pm}(t)$ is related to hypergeometric function \cite{Duverney:2024qzj,Michel2007}
	\begin{eqnarray}
		F^i_{\pm}(t)=
		{}_2F
		\left(
		\frac{\bar{\alpha}^i}{4}\mp\frac{\alpha^i}{4}+\frac{1}{2}, \ \frac{\bar{\alpha}^i}{4}\mp\frac{\alpha^i}{4}+\frac{1}{2}; \ 
		1\mp\frac{\alpha^i}{2}; \ 
		\frac{1-It}{2} 
		\right), 
		\label{Fpm-Scarf}
	\end{eqnarray}
	and $\alpha^i$ is a complex parameter
	\begin{eqnarray}
		\alpha^i={\sqrt{1-4(-1)^{i+1} kI}}.
		\label{alpha-bar}
	\end{eqnarray}
	Here we use $I$ to denote the imaginary unit (to avoid confusion with the index $i$). The calculation details are shown in Appendix. \ref{Cal-Hg}.
	
	Then substituting \eqref{t-U} into \eqref{two-Hg-t} we obtain the two linearly independent solutions of geodesic equation in \eqref{geo-scarf}
	\begin{eqnarray}
		X^{i}_{\pm}(U)=(\sinh U+I)^{\frac{1\mp\alpha^i}{4}} (\sinh U-I)^{\frac{1+\bar{\alpha}^i}{4}} \ F^i_{\pm}(\sinh U). 
		\label{two-Hg-U} 
	\end{eqnarray}
	
	\subsubsection{Asymptotic behavior of the test particle}
	
	Now let us study the asymptotic behavior of the test particle when $U\to\infty$. Note that $t\to\infty$ as $U\to\infty$, we will first study the asymptotic behavior of $X^i(t)$.
	
	Note that the hypergeometric function satisfies the following two identities \cite{Michel2007}
	\begin{eqnarray}
		&&
		{}_2F
		\left(
		a, \ 
		b; \ 
		c; \ 
		z 
		\right)=
		\frac{\Gamma(c)\Gamma(b-a)}{\Gamma(b)\Gamma(c-a)}
		(-z)^{-a} \ 
		{}_2F\left(
		a, \ 
		1-c+a; \ 
		1-b+a; \ 
		\frac{1}{z} 
		\right)
		\nonumber \\[8pt]
		&&\qquad\qquad\qquad\quad
		+\frac{\Gamma(c)\Gamma(a-b)}{\Gamma(a)\Gamma(c-b)}
		(-z)^{-b} \ 
		{}_2F\left(
		b, \ 
		1-c+b; \ 
		1-a+b; \ 
		\frac{1}{z} 
		\right),  \label{rela-Hg-1} \\[8pt]
		&&
		\lim_{z\to0}{}_2F
		\left(
		\left[
		a, \ 
		b
		\right]; \ 
		\left[
		c
		\right]; \ 
		z 
		\right)
		=1, \quad
		a,b,c\neq0,
		\label{rela-Hg-2}
	\end{eqnarray} 
	the first identity rewrites the hypergeometric function in terms of the inverse argument $1/z$, and is therefore suitable for studying the limit $z\to\infty$; then the latter allows the transformed hypergeometric functions appearing in \eqref{rela-Hg-1} to be reduced to constant values. Thus, these two identities are the basic ingredients needed to extract the asymptotic behavior of the solutions \eqref{two-Hg-t} when $t\to\infty$.
	
	However, there is an important subtlety in applying this procedure to the present Scarf case. Comparing \eqref{Fpm-Scarf} with the parameters in \eqref{rela-Hg-1}, one finds that the hypergeometric function is in the degenerate case $a=b$, the coefficients containing $\Gamma(b-a)$ and $\Gamma(a-b)$ are separately singular \footnote{The Gamma function $\Gamma(x)$ has a pole at $x=0$.}.
	
	To treat this case carefully, we temporarily split the two parameters by introducing a small regulator $\epsilon=b-a$, apply the connection formula \eqref{rela-Hg-1} for $\epsilon\neq0$, expand the contributions in $\epsilon$, and finally take the limit $\epsilon\to0$. This limiting procedure converts the pair of coincident power-law terms into a logarithmic asymptotic behavior. The detailed calculation is given in Appendix \ref{Cal-SAF}. The resulting asymptotic behavior is 
	\begin{eqnarray}
		\lim_{t\to\infty}X^i_{\pm}(t)\sim
		A_{\pm}^i+B_{\pm}^i\ln t, 
		\label{lim-t-X}
	\end{eqnarray}
	with the coefficients $A^i_\pm$ and $B^i_\pm$ given in  
	\begin{eqnarray}
		A_{\pm}^i=B_{\pm}^i\left(\mathcal{D}^i_{\pm}+\frac{\pi}{2}I-\ln2\right), \quad 
		B_{\pm}^i=\frac{\Gamma(c^i_{\pm})}{\Gamma(a^i_{\pm})\Gamma(c^i_{\pm}-a^i_{\pm})},
	\end{eqnarray}
	where
	\begin{eqnarray}
		\mathcal{D}^i_{\pm}=2\psi(1)-\psi(a^i_{\pm})-\psi(c^i_{\pm}-a^i_{\pm}), \quad \psi(x)=\frac{\dot{\Gamma}(x)}{\Gamma(x)},
		\label{D-psi-func}
	\end{eqnarray}
	the corresponding parameters are defined as
	\begin{eqnarray}
		a^i_{\pm}=b^i_{\pm}=\frac{1}{2}+\frac{\bar{\alpha}^i\mp\alpha^i}{4}, \quad c^i_{\pm}=1\mp\frac{\alpha^i}{2}.
		\label{para-X1}
	\end{eqnarray}
	
	The relation \eqref{t-U} tells us that the asymptotic behavior of $t$ is $t\sim e^{U}/2$ when $U\gg1$, then we finally obtain the asymptotic behavior of $X^i_{\pm}(U)$
	\begin{eqnarray}
		\lim_{U\to\infty}X^i(U)\sim C^i+D^iU,
		\label{lim-X-U-1}
	\end{eqnarray}
	where
	\begin{eqnarray}
		C^i=c_1A^i_++c_2A^i_-, \quad
		D^i=c_1B^i_++c_2B^i_-.
		\label{lim-X-U-2}
	\end{eqnarray}
	This means that the asymptotic behavior of the particle motion in Scarf profile is indeed the free motion.
	
	\subsubsection{Displacement memory effect in Scarf profile}
	
	We have so far demonstrated that, in the Scarf profile, the particle motion is strictly free. Nevertheless, a subtle distinction remains in the asymptotic behaviors of the two linearly independent solutions, $X^i_+$ and $X^i_-$.
	
	Note that the Gamma function $\Gamma(x)$ diverges whenever its argument is a negative integer \cite{DLMF}
	\begin{eqnarray}
		\Gamma(-n)=\infty, \quad \mathrm{when} \quad n\in\mathbb{Z}_{\geq0}.
		\label{Gamma-nn}
	\end{eqnarray}
	This implies that, with certain critical values of the amplitude $k$ and a suitable parameter selection, the velocity $D^i$ in \eqref{lim-X-U-2} may vanish, meaning that the asymptotic behavior of particle motion becomes a displacement memory effect with vanishing outgoing velocity. The presence or absence of this displacement motion becomes more evident in the two linearly independent solutions, $X^i_+$ and $X^i_-$.
		
	When $c_1=1$ and $c_2=0$, we have only $X^i_+$, whose asymptotic behavior can be written as 
	\begin{eqnarray}
			\lim_{t\to\infty}X^i_{+}(U)\sim
			A_{+}^i+B_{+}^iU, 
	\end{eqnarray}
	where
	\begin{eqnarray}
		A_{+}^i=B_{+}^i\left(\mathcal{D}^i_{+}+\frac{\pi}{2}I-\ln2\right), \quad 
		B_{+}^i=\frac{\Gamma(c^i_{+})}{\Gamma(a^i_{+})\Gamma(c^i_{+}-a^i_{+})}.
		\label{DM-X1-AB}
	\end{eqnarray}
	
	From \eqref{para-X1} and \eqref{alpha-bar} we know $c^i_{+}-a^i_{+}$ is a real number
	\begin{eqnarray}
		c^i_{+}-a^i_{+}=\frac{1-\Re(\alpha^i)}{2}. 
	\end{eqnarray}
	Then by assuming $c^i_{+}-a^i_{+}$ is a negative integer
	\begin{eqnarray}
		c^i_{+}-a^i_{+}=\frac{1-\Re(\alpha^i)}{2}=-n, \quad n\in\mathbb{Z}_{\geq0},
		\label{c-a-n}
	\end{eqnarray}
	we know that the Gamma function in $B^i_{+}$ develops a singularity
	\begin{eqnarray}
		\Gamma(c^i_{+}-a^i_{+})=\infty,
	\end{eqnarray}
	which leads to $B^i_{+}=0$, i.e., the outgoing velocity vanishes.
	
	On the other hand, $B^i_{+}=0$ does not imply that $A^i_+$ also vanishes, because in the expression \eqref{DM-X1-AB} for $A^i_+$ there is the term $\mathcal{D}^i_+$, and the function $\psi(c^i_+-a^i_+)$ in $\mathcal{D}^i_+$ diverges also when $c^i_{+}-a^i_{+}=-n$. The two divergences precisely cancel each other, resulting in a finite value of $A^i_+$. The calculation details can be found in Appendix. \ref{cal-DM-Scarf}. 
	
	The fact that $A^i_+$ remains finite when $c^i_+-a^i_+=-n$ indicates that although $X^i_+$ has no outgoing velocity, it possesses a finite displacement. In other words, the trajectory of $X^i_+$ corresponds to the displacement memory effect.
	
	Furthermore, from the relation \eqref{c-a-n} we can also obtain an expression for the critical value of the amplitude $k$:
	\begin{eqnarray}
		k^{\mathrm{cri}}_n=(1+2n)\sqrt{n(n+1)}, \quad n\in\mathbb{Z}_{\geq0}.
		\label{k-n}
	\end{eqnarray}
	This means that, when $k$ takes these discrete critical values, the trajectory of $X^i_+$ corresponds to a displacement memory effect; for other values of $k$, the trajectory corresponds to a velocity memory effect.
	
	Interestingly, Ref. \cite{Zhang:2025lxs} employed the Nikiforov-Uvarov method to solve \eqref{eq-S-t} and obtained the same critical relation \eqref{k-n} for the occurrence of the displacement memory effect.
	
	On the other hand, when $c_1=0$ and $c_2=1$ we have only $X^i_-$. However, the parameter ``$c^i_- - a^i_-$'' cannot be a negative integer since it is complex:
	\begin{equation}
		c^i_- - a^i_- = \frac{1 + I\cdot\Im(\alpha^i)}{2}.
	\end{equation}
	Consequently, only free motion escaping to infinity is possible for $X^i_-$.
	
\begin{figure}[htbp]
		\centering
		\subfloat[\label{ScarfX12}]{\includegraphics[width=0.5\linewidth]{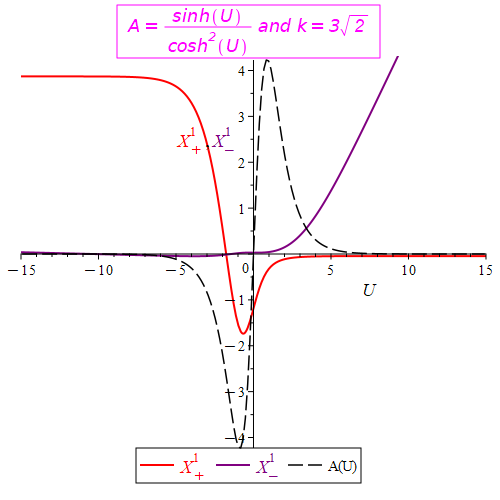}}
		\hfill
		\subfloat[\label{Xpm-1}]{\includegraphics[width=0.5\linewidth]{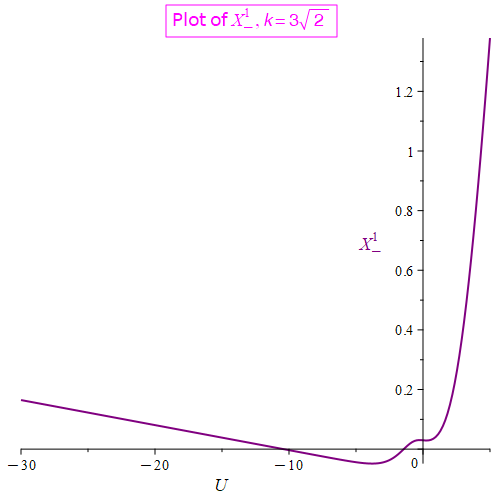}}
		\caption{
			\textit{\small 
				FIG. \ref{ScarfX12} shows the curves of two independent solutions when $n=1$. The \red{red curve} represents \red{$X_{+}^{1}$} corresponding to displacement motion, and the \purple{purple curve} represents the other independent solution \purple{$X_{-}^{1}$}. FIG. \ref{Xpm-1}	shows the curve of \purple{$X_{-}^{1}$} alone, the curve does not approach a constant in the outgoing region, but grows asymptotically.
			}
		}
		\label{Scarf-1}
	\end{figure}

	The two different asymptotic behaviors of $X^i_{+}$ and $X^i_-$ are shown in FIG. \ref{Scarf-1}, with an example for $k=3\sqrt{2}$ when $n=1$.
	
	\subsection{Weakly asymptotically free motion}\label{Weak-free}
	
	For weakly asymptotically free case, we take the critical behavior,
	\begin{eqnarray}
		A(U)=\frac{U}{(U^2+1)^2},
		\label{W-afm}
	\end{eqnarray}
	as a solvable case to study the details of the asymptotic motion of the test particle. Such profile exhibits the decay behavior $A(U)\sim U^{-3}$ when $U\gg1$, hence it satisfies the integral condition of weakly asymptotically free case in \eqref{A-cond-2}. For convenience, in the following we call this profile ``the inverse-cubic profile''.
	
\begin{figure}[htbp]
		\centering
		\includegraphics[scale=.5]{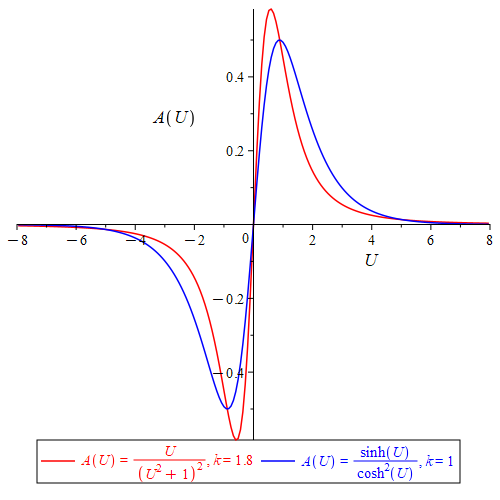}
		\caption{
			\textit{\small 
				This plot shows the shapes of the profiles in \eqref{W-afm} and in \eqref{Scarf}, they are very similar.
			}
		}
		\label{AU3-Scarf-1}
	\end{figure}

	Interestingly, the inverse-cubic profile in \eqref{W-afm} and the Scarf profile in \eqref{Scarf} share a remarkably similar shape, as shown in FIG. \ref{AU3-Scarf-1}. However, owing to their different tail behaviors, the asymptotic motions of the test particle differ.
	
	\subsubsection{Analytical solution of the motion of a test particle in the inverse-cubic profile}
	
	Under the inverse-cubic profile in \eqref{W-afm}, the transverse geodesic equation can be written as
	\begin{eqnarray}
		\frac{d^2}{dU^2}X^i(U)+(-1)^{i+1}k\frac{U}{(U^2+1)^2}X^i(U)=0,
	\end{eqnarray}
	whose analytical solution can also be obtained by using hypergeometric function
	\begin{eqnarray}
		X^i_{\pm}=c_1 \ X^i_++c_2 \ X^i_-,
		\label{U3-2-sol}
	\end{eqnarray}
	where
	\begin{eqnarray}
		&&
		X^i_{\pm}=
		(I+U)^{\frac{1\mp\bar{\beta}^i}{2}}
		(I-U)^{\frac{1+{\beta}^i}{2}}
		F^i_{\pm}(U), \label{U3-Xpm}
		\\[8pt] 
		&&
		F^i_{\pm}(U)=
		{}_2F\left(\frac{\beta^i\mp\bar{\beta}^i}{2},1+\frac{\beta^i\mp\bar{\beta}^i}{2};1\mp\bar{\beta}^i;\frac{1-IU}{2}\right),
	\end{eqnarray}
	with
	\begin{eqnarray}
		\beta^i=\sqrt{1+(-1)^{i+1}kI}.
		\label{U3-beta}
	\end{eqnarray}
	
	\subsubsection{Asymptotic behavior of the test particle in the inverse-cubic profile}
	
	Now let us study the asymptotic behavior of the test particle when $U\to\infty$.
	
	For convenience we first set $z=\frac{1-IU}{2}$. Then, the two linearly independent solutions in \eqref{U3-Xpm} can be written as
	\begin{eqnarray}
		X^i_{\pm}(z)=
		(2-2z)^{\frac{1\mp\bar{\beta}^i}{2}}
		(2z)^{\frac{1+{\beta}^i}{2}}
		F^i_{\pm}(z).
		\label{U3-Xpm-z}
	\end{eqnarray}
	
	By using again the two identities of hypergeometric function in \eqref{rela-Hg-1}-\eqref{rela-Hg-2}, we obtain the asymptotic expression of the above when $z\to-\infty$
	\begin{eqnarray}
		\lim_{z\to-\infty}X^i_{\pm}(z)
		\sim
		\mathcal{A}^{i}_{\pm}\left[(-z)-p^i\ln(-z)\right]+M^i_{\pm},
		\label{lim-U3-Xpm-z-final}
	\end{eqnarray}
	where
	\begin{eqnarray}
		\mathcal{A}^i_{\pm}=\frac{\Gamma(\tilde{c}^i_{\pm})}{\Gamma(\tilde{b}^i_{\pm})\Gamma(\tilde{c}^i_{\pm}-\tilde{a}^i_{\pm})}, \quad
		p^i=\frac{(-1)^{i}kI}{2},
		\label{zw-AB}
	\end{eqnarray}
	with
	\begin{eqnarray}
		\tilde{a}^i_{\pm}=\frac{\beta^i\mp\bar{\beta}^i}{2}, \quad
		\tilde{b}^i_{\pm}=1+\frac{\beta^i\mp\bar{\beta}^i}{2}, \quad
		\tilde{c}^i_{\pm}=1\mp\bar{\beta}^i,
	\end{eqnarray}
	and $M^i_{\pm}$ denotes the finite part of the asymptotic expansion after subtracting the leading linear and logarithmic terms. It is determined by the hypergeometric connection coefficients and by the branch choice of $X^i_{\pm}$, and is not an additional integration constant. The calculation details can be found in Appendix. \ref{Cal-WAF}.
	
	Substituting \eqref{lim-U3-Xpm-z-final} into \eqref{U3-2-sol} and considering $z=\frac{1-IU}{2}$, we finally obtain the asymptotic behavior of the test particle
	\begin{eqnarray}
		\lim_{U\to\infty}X^i(U)\sim\tilde{M}^i+\mathcal{V}^i\cdot\left[U-(-1)^{i}k\ln U\right], 
		\label{U3-asym-motion}
	\end{eqnarray}
	where $\tilde{M}^i$ is a constant originating from $M^i_{\pm}$ and the other parameters, and $\mathcal{V}^i$ is the leading velocity
	\begin{eqnarray}
		\mathcal{V}^i=c_1\frac{\mathcal{A}^i_{+}I}{2}+c_2\frac{\mathcal{A}^i_{-}I}{2}.
		\label{U3-Vi}
	\end{eqnarray}
	
\begin{figure}[htbp]
		\centering
		\subfloat[\label{U3-X-free}]{\includegraphics[width=0.49\linewidth]{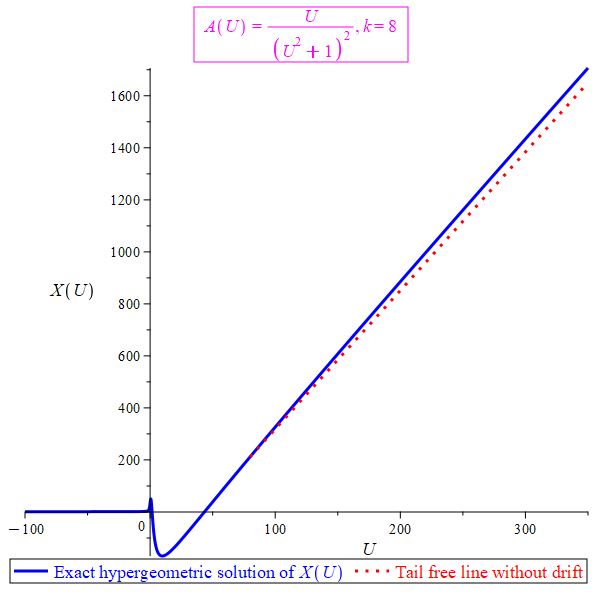}}
		\hfill
		\subfloat[\label{U3-tail-X-free}]{\includegraphics[width=0.49\linewidth]{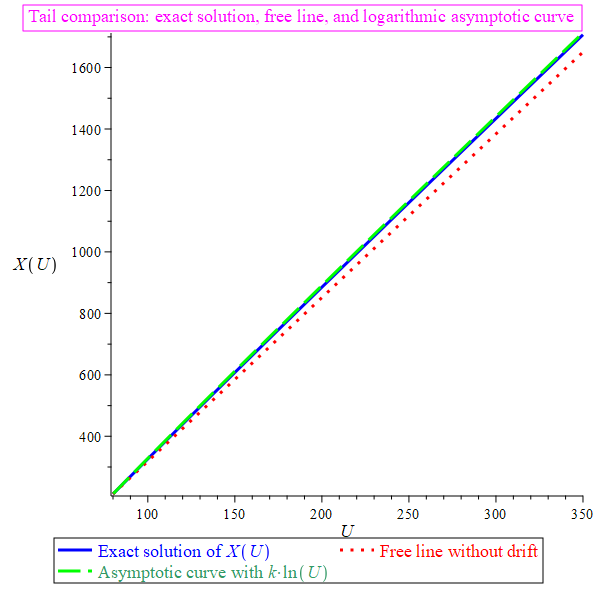}}
		\caption{
			\textit{\small 
				FIG. \ref{U3-X-free} compares the exact solution obtained from the hypergeometric function (\blue{solid blue line}) with the drift-free free motion (\red{dotted red line}). The departure between their asymptotic tails arises from the drift term ``$k\ln U$''. FIG. \ref{U3-tail-X-free} displays the comparison among the exact hypergeometric solution $X(U)$ (\blue{blue solid curve}), the asymptotic solution in \eqref{U3-asym-motion} (\green{green dashed line}), and the drift-free free motion (\red{red dotted line}). The $\ln U$ drift in the asymptotic tail is shown even more clearly.
			}
		}
		\label{U3-drift}
	\end{figure}

	From \eqref{U3-asym-motion}, we see that for the inverse-cubic profile, whose tail behaves as $A(U)|_{U\gg1}\sim U^{-3}$, the asymptotic motion of a test particle is not purely free motion, but free motion accompanied by a drift term ``$k\ln U$''. FIG. \ref{U3-drift} illustrates this asymptotic motion with a drift.
	
	Notice that the drift term ``$k\ln U$'' in \eqref{U3-asym-motion} is tied to $\mathcal{V}^i$ also, therefore, whether a drift appears depends on the presence of the leading velocity term ``$\mathcal{V}^iU$''. Interestingly, even with a drift term, the true velocity of the test particle remains finite as $U\to\infty$,
	\begin{eqnarray}
		\lim_{U\to\infty}V^i(U)=\lim_{U\to\infty}\left(\frac{dX^i}{dU}\right)=\lim_{U\to\infty}\left(\mathcal{V}^i-(-1)^{i}\frac{k}{U}\right)=\mathcal{V}^i<\infty.
	\end{eqnarray}
	The effect of the drift therefore manifests itself only in the free motion governed by the leading velocity term ``$\mathcal{V}^iU$'': whenever there is a leading velocity, a drift necessarily accompanies it. This precisely corresponds to the integral condition in \eqref{A-cond-2}: the integral for the outgoing velocity converges, giving a leading velocity term, while the tail integral diverges, giving rise to a divergent intercept and a drift term.
	
	However, when the outgoing velocity vanishes, this drift term is also absent, and then the residual effect is still a displacement memory, as discussed below.
	
	\subsubsection{Displacement memory effect in the inverse-cubic profile}
	
	From \eqref{U3-Vi} and \eqref{zw-AB} we know that the leading velocity is related to the Gamma function
	\begin{eqnarray}
		\mathcal{V}^i\propto \frac{1}{\Gamma(\tilde{c}^i_{\pm}-\tilde{a}^i_{\pm})},
	\end{eqnarray}
	This means that we can still exploit the singularity of the Gamma function at negative integer arguments to make the outgoing velocity vanish, while the finite part $\tilde{M}^i_{\pm}$ is generally retained, and the corresponding solution then exhibits the displacement memory effect. 
	
	Notice that $\tilde{c}^i_+-\tilde{a}^i_+$ is real, whereas $\tilde{c}^i_--\tilde{a}^i_-$ is complex,
	\begin{eqnarray}
		\tilde{c}^i_+-\tilde{a}^i_+=1-\Re(\beta^i), \quad
		\tilde{c}^i_--\tilde{a}^i_-=1-\Im(\beta^i)\cdot I.
	\end{eqnarray}
	Hence, only the linearly independent solution $X^i_+$ can give rise to the displacement memory effect. This is consistent with the case of strongly asymptotically free motion.
	
\begin{figure}[htbp]
		\centering
		\subfloat[\label{U3-DM-1}]{\includegraphics[width=0.49\linewidth]{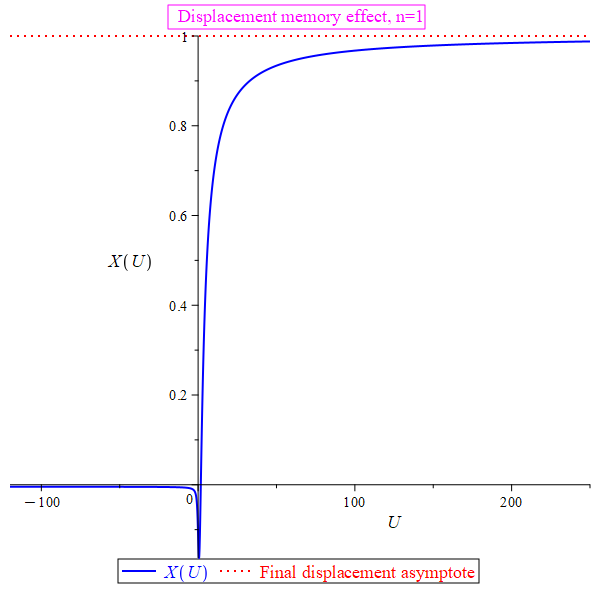}}
		\hfill
		\subfloat[\label{U3-DM-2}]{\includegraphics[width=0.49\linewidth]{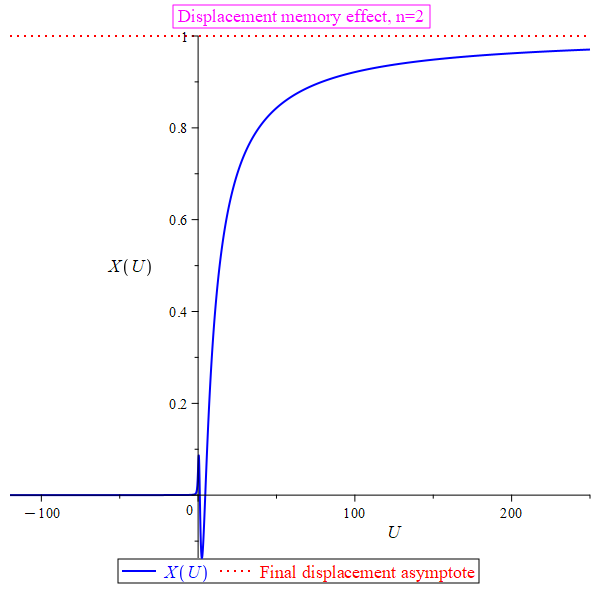}}
		\caption{
			\textit{\small 
				FIG. \ref{U3-DM-1} and \ref{U3-DM-2} illustrate the displacement memory effect for $n=1$ and $n=2$, respectively, with \blue{the blue solid curves} showing the motion and \red{the red dotted curves} marking the asymptotic position.
			}
		}
		\label{U3-DM}
	\end{figure}

	When $\tilde{c}_+-\tilde{a}_+$ is a negative integer $\tilde{c}_+-\tilde{a}_+=-n$, we can still obtain the expression for the critical values of $k$,
	\begin{eqnarray}
		k^{\mathrm{cri}}_n=2(n+1)\sqrt{n(n+2)}, \quad n\in\mathbb{Z}_{\geq0}.
	\end{eqnarray}
	These critical values remain discrete. The corresponding displacement-memory motions for $n=1,2$ are illustrated in FIG. \ref{U3-DM}.
	
	\subsection{Non-asymptotically free motion}
	
	For non-asymptotically free motion, we take the critical behavior,
	\begin{eqnarray}
		A(U)=\frac{1}{U^2+1}, \label{A-U-2}
	\end{eqnarray}
	as a solvable case to study the details. For convenience, we call this profile ``inverse-square profile''. The asymptotic behavior of such profile is $A(U)|_{U\gg1}\sim U^{-2}$; hence it satisfies the condition of non-asymptotically free case in \eqref{A-cond-3}.
	
\begin{figure}[htbp]
		\centering
		\includegraphics[scale=.55]{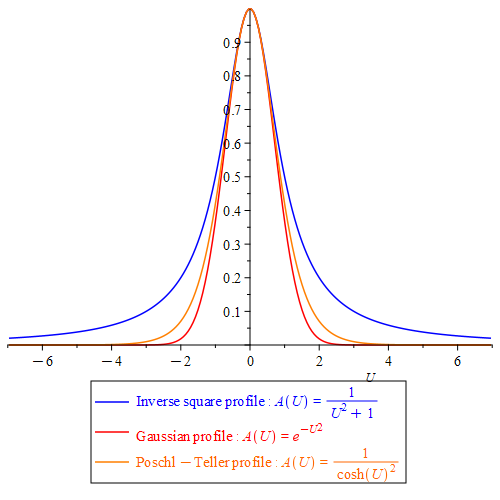}
		\caption{
			\textit{\small 
				This plot illustrates the similar shapes of the inverse-square profile, the Gaussian profile, and the P\"oschl–Teller profile. However, owing to their different asymptotic behaviors, the inverse-square profile decays more slowly, while the Gaussian and P\"oschl-Teller profiles fall off more rapidly.
			}
		}
		\label{U2-GP-PT}
	\end{figure}

	The shape of this profile is similar to those of the P\"oschl-Teller and Gaussian profiles, as shown in FIG. \ref{U2-GP-PT}. 
	
	However, despite the similarity in shape, the motion of a test particle within these profiles differs significantly. The P\"oschl-Teller and Gaussian profiles exhibit exponential asymptotic behavior and therefore satisfy \eqref{A-cond-1}; consequently, the test particle undergoes strictly asymptotically free motion. Relevant details can be found in Refs.  \cite{Zhang:2024uyp,Elbistan:2025wbp}. In contrast, the inverse-square profile decays as $A(U)|_{U\to\infty}\sim U^{-2}$, which is much slower than the P\"oschl-Teller or Gaussian profile, and satisfies only \eqref{A-cond-3}. Its asymptotic motion is therefore drastically different from the cases discussed earlier, as will be shown in the calculations that follow.

	\subsubsection{Analytical solution of the motion of a test particle}
	
	The transverse geodesic equation under the inverse-square profile \eqref{A-U-2} can be written as
	\begin{eqnarray}
		\frac{d^2}{dU^2}X^i(U)+(-1)^{i+1}\frac{k}{U^2+1}X^i(U)=0.
		\label{eq-geo-u2}
	\end{eqnarray}
	The solutions of the above equation can also be obtained by using hypergeometric equation
	\begin{eqnarray}
		X^i(U)=c_+ \ (U^2+1)F^i_{+}(U)+c_- \ U(U^2+1)F^i_{-}(U),
		\label{anal-X-1}
	\end{eqnarray}
	where
	\begin{eqnarray}
		F^i_{\pm}(U)={}_2F\left(1\mp\frac{1+\rho^i}{4},1\mp\frac{1-\rho^i}{4};1\mp\frac{1}{2};-U^2\right),
		\label{anal-X-2}
	\end{eqnarray}
	with
	\begin{eqnarray}
		\rho^i=\sqrt{1-4(-1)^{i+1}k}.
		\label{anal-X-3}
	\end{eqnarray}
	
	\subsubsection{Asymptotic behavior of test particle with different $k$}
	
	By employing \eqref{rela-Hg-1}-\eqref{rela-Hg-2} and using the same technique described in Appendix. \ref{Cal-SAF} and \ref{Cal-WAF}, we can directly obtain the final results of asymptotic behavior, as shown in following:
	\begin{eqnarray}
		&&
		\lim_{U\to\infty}X^i(U)\sim
		\left\{
		\begin{array}{ccc}
			C^i_+ \ U^{\frac{1+\rho^i}{2}}+C^i_- \ U^{\frac{1-\rho^i}{2}}, \quad &(-1)^{i+1}k<\frac{1}{4} \\
			U^{\frac{1}{2}}\left( D^i_++D^i_-\ln U \right),
			\quad &(-1)^{i+1}k=\frac{1}{4} \\
			U^{\frac{1}{2}}\left[E^i_+\cos\left(\frac{|\rho^i|}{2}\ln U\right)+IE^i_-\sin\left(\frac{|\rho^i|}{2}\ln U\right)\right], \quad &(-1)^{i+1}k>\frac{1}{4}
		\end{array}
		\right..
		\label{U2-asym-X}
	\end{eqnarray}
	where
	\begin{eqnarray}
		&&
		C^i_{\pm}=
		c_{\pm}\mathfrak{A}^i_{\pm}\Gamma(m^i_{\pm}-l^i_{\pm})
		+
		c_{\mp}\mathfrak{B}^i_{\mp}\Gamma(l^i_{\mp}-m^i_{\mp}), 
		\\[8pt]
		&&
		D^i_+=2\dot{\Gamma}(1) 
		\cdot
		\left(
		c_-\mathfrak{A}^i_{-}
		+
		c_+\mathfrak{A}^i_{+}
		\right),
		\quad
		D^i_-=2
		\left(
		c_-\mathfrak{A}^i_{-}
		-
		c_+\mathfrak{A}^i_{+}
		\right),
		\\[8pt]
		&&
		E^i_{\pm}=
		c_+
		\left[
		\mathfrak{A}^i_{+}\Gamma(m^i_{+}-l^i_{+})
		\pm
		\mathfrak{B}^i_{+}\Gamma(l^i_{+}-m^i_{+})
		\right]
		\nonumber \\
		&&\qquad
		+
		c_-
		\left[
		\mathfrak{B}^i_{-}\Gamma(l^i_{-}-m^i_{-})
		\pm
		\mathfrak{A}^i_{-}\Gamma(m^i_{-}-l^i_{-})
		\right],
	\end{eqnarray}
	with
	\begin{eqnarray}
		\mathfrak{A}^i_{\pm}=\frac{\Gamma(n^i_{\pm})}{\Gamma(m^i_{\pm})\Gamma(n^i_{\pm}-l^i_{\pm})}, \quad
		\mathfrak{B}^i_{\pm}=\frac{\Gamma(n^i_{\pm})}{\Gamma(l^i_{\pm})\Gamma(n^i_{\pm}-m^i_{\pm})},
	\end{eqnarray}
	and 
	\begin{eqnarray}
		l^i_{\pm}=1\mp\frac{1+\rho^i}{4}, \quad
		m^i_{\pm}=1\mp\frac{1-\rho^i}{4}, \quad
		n^i_{\pm}=1\mp\frac{1}{2}.
	\end{eqnarray}
	The asymptotic motion of a test particle behaves distinctly for different $k$. 

\begin{figure}[htbp]
		\centering
		\subfloat[\label{U2-X-1}]{\includegraphics[width=0.33\linewidth]{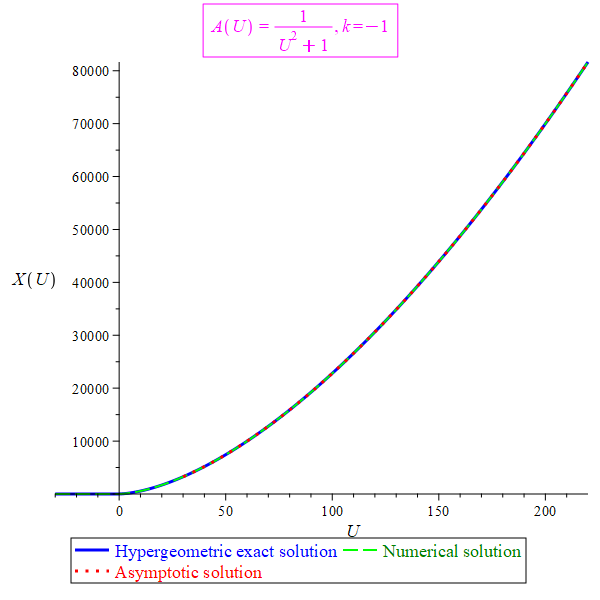}}
		\hfill
		\subfloat[\label{U2-X-2}]{\includegraphics[width=0.33\linewidth]{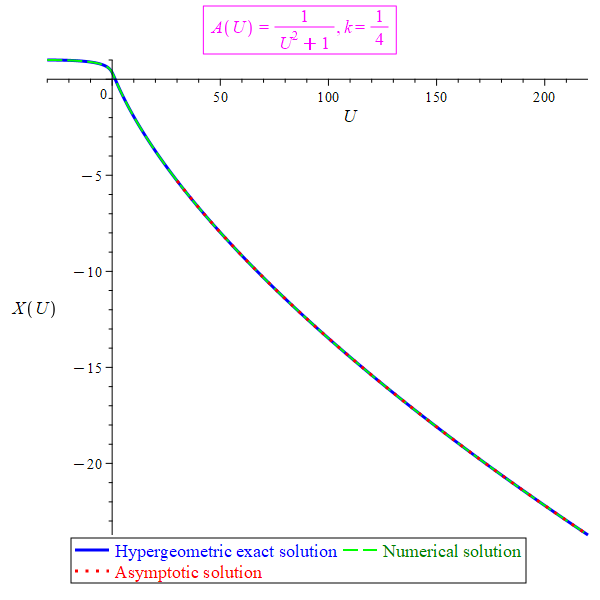}}
		\hfill
		\subfloat[\label{U2-X-3}]{\includegraphics[width=0.33\linewidth]{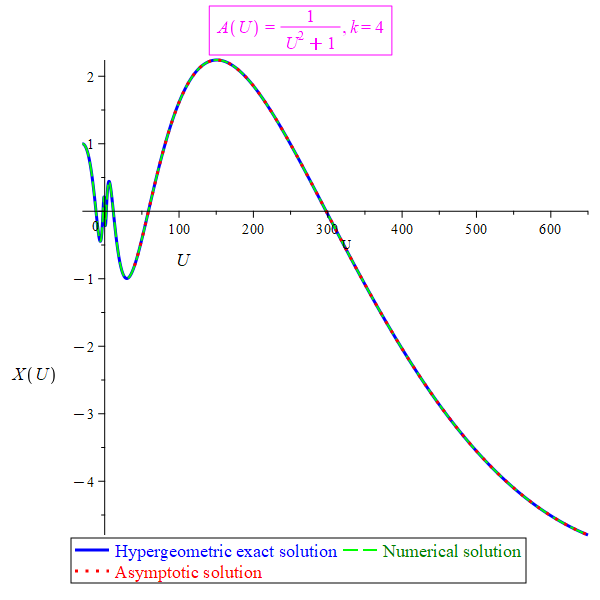}}
		\caption{
			\textit{\small  
				FIG. \ref{U2-X-1}, \ref{U2-X-2} and \ref{U2-X-3} display the motion for $k=-1$, $k=\frac{1}{4}$, and $k=4$, respectively, which correspond to the power-law solution, the composite power-law and logarithmic motion, and the logarithmically oscillatory motion with slowly growing amplitude. In each plot, \blue{the solid blue curve} represents the exact solution described by the hypergeometric function in \eqref{anal-X-1}-\eqref{anal-X-3}, \green{the green dashed curve} shows the numerical result, and \red{the red dotted curve} corresponds to the asymptotic result in \eqref{U2-asym-X}.
			}
		}
		\label{U2-X}
	\end{figure}

	From \eqref{U2-asym-X} we see that, regardless of the value of $k$, the asymptotic behavior of $X^i(U)$ is no longer free; in fact, it cannot even exhibit the weakly asymptotically free motion. The critical value $(-1)^{i+1}k=\frac{1}{4}$ divides the motion into three distinct types: 
	\begin{itemize}
		\item 
		when $(-1)^{i+1}k>\frac{1}{4}$, it is a logarithmically oscillatory motion with slowly growing amplitude;
		
		\item 
		when $(-1)^{i+1}k=\frac{1}{4}$, it is a composite motion of a $\frac{1}{2}$-power term and a logarithmic term;
		
		\item
		when $(-1)^{i+1}k<\frac{1}{4}$, it is a pure power-law motion.
	\end{itemize}
	The three plots in FIG. \ref{U2-X} illustrate the curves of these motions for $X(U)$, respectively.
	
	\section{An Intrinsic Physical Perspective: the Geodesic Deviation Equation and the Tidal Matrix}\label{Intrinsic Physical Perspective}
	
    In the previous sections we began with the motion of a single test particle and studied the transverse geodesic equations in Brinkmann coordinates. These results are useful since they directly display how the tail of the wave profile $A(U)$ affects the asymptotic motion. In particular, they show that the decay rates $A(U)\sim U^{-3}$ and $A(U)\sim U^{-2}$ represent two critical behaviors: the former separates strongly asymptotically free motion from weakly asymptotically free motion, while the latter marks the threshold of non-asymptotically free motion.
    
    However, the trajectory of a single particle should not be regarded as the final physical observable. The coordinate position, coordinate velocity, and even the coordinate notion of ``free motion'' may depend on the choice of coordinates. Therefore, while the geodesic analysis of a single particle provides a useful diagnostic, it is not by itself sufficient to define a coordinate-independent physical classification.
    
    To extract the intrinsically invariant property, we now turn to the relative motion of a family of neighbouring freely falling particles. This is also closer to the operational meaning of GW detection: a detector does not measure the coordinate trajectory of an isolated particle, but rather the relative displacement, relative velocity, and relative acceleration between different test masses.
	
	The separation vector $\xi^{\mu}$ between neighbouring geodesics satisfies the covariant geodesic-deviation equation
	\begin{eqnarray}
		&&
		\frac{D^2}{D\tau^2}\xi^{\mu}+R^{\mu}_{\  \alpha\nu\beta}u^{\alpha}\xi^{\nu}u^{\beta}=0. 
		\label{gd-eq}
	\end{eqnarray}
	For the vacuum plane GW in \eqref{metric-PGW}, the Ricci tensor vanishes, leading to the fact that the Riemann tensor coincides with the Weyl tensor
	\begin{eqnarray}
		R^{\mu}_{\  \alpha\nu\beta}=C^{\mu}_{\  \alpha\nu\beta}.
	\end{eqnarray}
	The non-vanishing components are listed in follow
	\begin{eqnarray}
		R^{V}_{\  iUj}=C^{V}_{\  iUj}=\delta_{ij}(-1)^{i+1}kA(U), \quad
		R^{i}_{\  UjU}=C^{i}_{\  UjU}=\delta^i_{ \ j}(-1)^{i+1}kA(U).
		\label{non0-Weyl}
	\end{eqnarray}
	
	By employing \eqref{non0-Weyl}, \eqref{gd-eq} gives us
	\begin{eqnarray}
		&&
		\frac{d^2}{dU^2}\xi^{i}+\mathcal{K}^i_{ \ j}\xi^{j}=0, \quad
		\mathcal{K}^i_{ \ j}=C^{i}_{\  UjU}=\delta^i_{ \ j}(-1)^{i+1}kA(U),
		\label{eq-xi-i} \\
		&&
		\frac{d^2}{dU^2}\xi^{V}-k\left(X\xi^{X}-Y\xi^{Y}\right)\frac{dA}{dU}-2kA(U) \frac{d}{dU}\left( X\xi^{X}-Y\xi^{Y} \right)=0, \label{eq-xi-V}
	\end{eqnarray}
	where we have already assumed $\xi^{U}=0$, since we are discussing the relative motion of test particles on the same wave-front, where they maintain simultaneity. 
	
	From \eqref{eq-xi-i}-\eqref{eq-xi-V} we can see that the free components in the geodesic deviation equation are still the $X$ and $Y$ components on the transverse plane; $\mathcal{K}^i_{ \ j}$ is the tidal matrix \cite{Carneiro:2026iuk,Flanagan:2019ezo}, which comes from the projection of the Weyl tensor onto the transverse wave-front. 
	
	Furthermore, we can find that the form of the geodesic-deviation equation for the transverse wave-front components in \eqref{eq-xi-i} is completely consistent with the geodesic equation for a single particle in \eqref{geo-X-i}, which implies that our previous classification of the profile possesses a deeper physical meaning: \emph{this classification is not merely of wave profiles, but more fundamentally of the tidal forces within the GW spacetime, and is itself an intrinsic physical property.} It divides the relative asymptotic motion of test particles into three types:
	\begin{itemize}
		\item 
		Strongly asymptotically free relative motion:
		\begin{eqnarray}
			\int_{U_0}^{\infty}s^2||\mathcal{K}(s)||ds <\infty;
		\end{eqnarray}
		
		\item 
		Weakly asymptotically free relative motion:
		\begin{eqnarray}
			\int_{U_0}^{\infty} s||\mathcal{K}(s)||ds <\infty, 
			\quad \mathrm{but} \quad
			\int_{U_0}^{\infty} s^2||\mathcal{K}(s)||ds \to\infty;
		\end{eqnarray}
		
		\item 
		Non-asymptotically free relative motion:
		\begin{eqnarray}
			\int_{U_0}^{\infty}s||\mathcal{K}(s)||ds \to\infty.
		\end{eqnarray}
		
	\end{itemize}
	Here the notation ``$||\cdot||$'' means the norm of $\mathcal{K}^i_j$. These norm conditions are robust sufficient criteria and do not include possible improvements due to oscillatory cancellations.
	
	Moreover, since the geodesic deviation equation is covariant, this classification can be formulated covariantly in terms of the tidal matrix along a reference geodesic, and is therefore not an artifact of Brinkmann coordinate trajectories.
	
	\section{Summary}\label{Summary}
	
	The condition $A(U)|_{U\to\infty}=0$ is analogous to saying that the profile for plane waves vanishes at infinity. However, this is not sufficient to guarantee ordinary free motion, the latter requires more stringent conditions. In this paper, we study how the tail of plane wave profiles affect the asymptotic motion of test particles. Based on the decay rate of profile, we classify the asymptotic motion of the particles into three types: strongly asymptotically free motion, weakly asymptotically free motion, and non-asymptotically free motion. This classification is not merely an artifact of a particular coordinate system for the single-particle motion; rather, it is an intrinsic property of the tidal forces of GWs, and applies to the relative motion of neighbouring test particles when formulated in terms of the tidal matrix along a reference geodesic.
	
	 What matters is the accumulated tail along the would-be free trajectory. In the plane-wave geodesic (deviation) equation, the tidal acceleration is $A(U)X^i(U)$ (or $\mathcal{K}^i_{ \ j}\xi^j$ for deviation equation), and for a generic free trajectory, the accumulated velocity and intercept involve the weighted terms $\int_{U_*}^{\infty}s|A(s)|ds$ and $\int_{U_*}^{\infty}s^2|A(s)|ds$. Therefore the distinction between short-range and long-range plane-wave tails is directly analogous to the familiar distinction between short-range and long-range forces in scattering theory \cite{Li:2021ekj}.
	
	A related recent development is the on-shell analysis of memory and supertranslations on plane-wave spacetimes by Cristofoli and Klisch \cite{Cristofoli:2025esy}. In their work, the gravitational velocity memory is encoded in the matrix $c_i^{a}$ appearing in the out-region asymptotics of the transverse vielbein, $E_i^a=b_i^a+x^-c_i^a$. Their focus is on how this memory data enters the exact gravitational waveform emitted by a massive test particle scattering on the plane-wave background, including contributions from Synge's world function, Green-function tail terms, and the dependence on the choice of BMS frame. By contrast, our work addresses a prior geometric question: for plane-wave profiles which are not necessarily sandwich but only satisfy $A(U)|_{U\to\infty}\to0$, under what decay conditions does such a free asymptotic expansion exist at all? In this sense, our weighted-integrability criteria classify when the velocity-memory data used in scattering descriptions remain asymptotically well defined for long-range plane-wave tails.
	
	The bulk velocity memory studied in this paper should be distinguished from the BMS-frame dependence of radiative waveforms at $\mathscr{I}^+$. In the Brinkmann description, the velocity-memory data are determined by the accumulated tidal evolution in the plane-wave bulk and do not require fixing a Bondi or BMS frame. In the on-shell analysis of Ref.~\cite{Cristofoli:2025esy}, the BMS-frame choice acts on the particle-emitted waveform at $\mathscr{I}^+$; the induced phase shift of the background waveshape is only a light-front coordinate shift. Such a shift leaves the velocity-memory data and the asymptotic class unchanged, up to the trivial redefinition of the free-line intercept under a translation of $U$.
	
	On the other hand, the amplitude-based analysis of Adamo, Cristofoli, Ilderton and Klisch computed all-order classical waveforms for scattering on gravitational plane waves \cite{Adamo:2022qci}, treating the plane wave as an exact nonlinear background. Their setup assumes a compactly supported sandwich profile in order to define an S-matrix, and for simplicity takes the induced velocity memory to be parametrically small. In that setting the standard free in/out regions (or before/after zones in our paper) are part of the construction. The later analysis of Cristofoli and Klisch relaxed this weak-memory approximation and showed how the full velocity-memory data enter the waveform and the BMS-frame dependence \cite{Cristofoli:2025esy}. Our work is complementary to both: rather than computing the waveform, we ask under what conditions on a non-compact plane-wave profile $A(U)|_{U\to\infty}=0$ the free asymptotic data used in such scattering descriptions exist in the first place.

	\begin{acknowledgments} \vskip-3mm
		We deeply appreciate Professor Rong-Gen Cai for his constant encouragement and support. This work is supported in part by the National Natural Science Foundation of China with grants No. 12475063, No. 12075232.
	\end{acknowledgments}
	\goodbreak

	\appendix
	\section{General integral solution of geodesic equation}\label{Cal-Int-sol}
	
	Now we show the calculation details about how to obtain \eqref{geo-sol-i}.
	
	We first take the first integral on Eq. \eqref{geo-X-i}, then obtain
	\begin{eqnarray}
		{X^{i}}'=\mathcal{V}^i_0+(-1)^{i} \ k\int_{U_0}^{U}ds \ A(s)X^i(s),
	\end{eqnarray}
	where $(\cdot)'$ means $d/dU$, $\mathcal{V}^i_0={X^{i}}'(U_0)$ means the initial  velocity of the test particle.
	
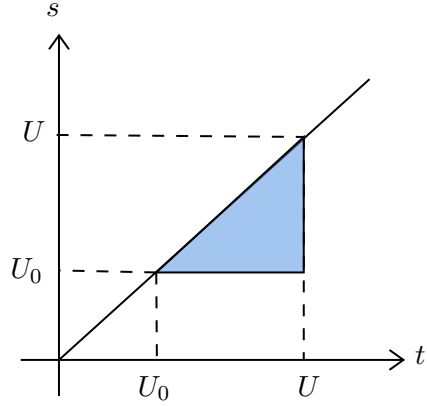
\begin{figure}[htbp]
	 	\centering

	 	\tikzset{every picture/.style={line width=0.75pt}} %set default line width to 0.75pt        
	 	
	 	\begin{tikzpicture}[x=0.75pt,y=0.75pt,yscale=-1,xscale=1]
	 		
	 		%Shape: Axis 2D [id:dp07274944223990731] 
	 		\draw  (226.67,212.82) -- (418.67,212.82)(245.87,49.92) -- (245.87,230.92) (411.67,207.82) -- (418.67,212.82) -- (411.67,217.82) (240.87,56.92) -- (245.87,49.92) -- (250.87,56.92)  ;
	 		%Straight Lines [id:da4853965577841781] 
	 		\draw    (401.67,71.92) -- (245.87,212.82) ;
	 		%Straight Lines [id:da5930527833515626] 
	 		\draw  [dash pattern={on 4.5pt off 4.5pt}]  (294.67,168.92) -- (295,213) ;
	 		%Straight Lines [id:da4184898927888523] 
	 		\draw  [dash pattern={on 4.5pt off 4.5pt}]  (368.67,100.92) -- (368.67,212.92) ;
	 		%Straight Lines [id:da31571364721328066] 
	 		\draw  [dash pattern={on 4.5pt off 4.5pt}]  (294.67,168.92) -- (245.67,167.92) ;
	 		%Straight Lines [id:da34320109453109304] 
	 		\draw  [dash pattern={on 4.5pt off 4.5pt}]  (244.67,99.92) -- (368.67,100.92) ;
	 		%Shape: Right Triangle [id:dp513798245273595] 
	 		\draw  [fill={rgb, 255:red, 74; green, 144; blue, 226 }  ,fill opacity=0.51 ] (368.67,100.92) -- (294.67,168.92) -- (368.67,168.92) -- cycle ;
	 		
	 		% Text Node
	 		\draw (238,32.4) node [anchor=north west][inner sep=0.75pt]    {$s$};
	 		% Text Node
	 		\draw (423,204.4) node [anchor=north west][inner sep=0.75pt]    {$t$};
	 		% Text Node
	 		\draw (284,219.4) node [anchor=north west][inner sep=0.75pt]    {$U_{0}$};
	 		% Text Node
	 		\draw (364,219.4) node [anchor=north west][inner sep=0.75pt]    {$U$};
	 		% Text Node
	 		\draw (220,160.4) node [anchor=north west][inner sep=0.75pt]    {$U_{0}$};
	 		% Text Node
	 		\draw (226,91.4) node [anchor=north west][inner sep=0.75pt]    {$U$};

	 	\end{tikzpicture}
	 	
	 	\caption{This plot shows the double integral in \eqref{o-double-int}. \label{double-int-plot}}
	 	
	 \end{figure}

	Then we take the second integral, and obtain
	\begin{eqnarray}
		X^{i}=X^{i}_{0}+\mathcal{V}^i_0(U-U_0)+(-1)^i \ k\int_{U_0}^{U}dt\int_{U_0}^{t}ds \ A(s)X^i(s).
		\label{o-double-int}
	\end{eqnarray}
	The integral on the right hand side in \eqref{o-double-int} can be viewed as a double integral over a two-dimensional triangle, as shown in FIG. \ref{double-int-plot}.
	
	The order of integration is first to integrate $s$ from $U_0$ to $t$, and then to integrate $t$ from $U_0$ to $U$. So the integration first sweeps along the vertical $s$-axis, then along the horizontal $t$-axis. By swapping this order of integration, we have 
	\begin{eqnarray}
		&&
		X^i=X^i_0+\mathcal{V}^i_0(U-U_0)+(-1)^i \ k\int_{U_0}^{U}ds\int_{s}^{U}dt \ A(s)X^i(s) \nonumber \\
		&&
		\qquad\quad
		=X^i_0+\mathcal{V}^i_0(U-U_0)+(-1)^i \ k\int_{U_0}^{U}ds \ (U-s)A(s)X^i(s).
	\end{eqnarray}
	
	For the test particle initially at rest, the initial velocity should vanish $\mathcal{V}^i_{0}=0$, then we recover \eqref{geo-sol-i}.
	
	\section{Calculations for hypergeometric solution}\label{Cal-Hg}
	
	\eqref{eq-S-t} has three regular singularities at $t=\pm I,\infty$, therefore it can be transformed to Gauss hypergeometric equation.
	
	Setting 
	\begin{eqnarray}
		z=\frac{1-It}{2},
	\end{eqnarray}
	the three regular singularities can be moved to $z=0,1,\infty$, then \eqref{eq-S-t} becomes
	\begin{eqnarray}
		z(1-z)\frac{d^2}{dz^2}X^i(z)+\left(\frac{1}{2}-z\right)\frac{d}{dz}X^i(z)+\frac{I(-1)^{i+1}k(1-2z)}{4z(1-z)}X^i(z)=0.
		\label{X-z-eq}
	\end{eqnarray}
	
	Then using the following transformation 
	\begin{eqnarray}
		X^i(z)\to z^{\rho}(1-z)^{\sigma}X^i(z),
	\end{eqnarray}
	\eqref{X-z-eq} can be written as
	\begin{eqnarray}
		&&
		z(1-z)\frac{d^2}{dz^2}X^i(z)+\left(\frac{1}{2}-(2\rho + 2\sigma+1)z+2\rho\right)\frac{d}{dz}X^i(z) 
		\nonumber \\
		&&
		\quad
		+\left[\frac{\rho^2 - \frac{\rho}{2} + \frac{(-1)^{i+1}kI}{4}}{z} + \frac{\sigma^2 - \frac{\sigma}{2} - \frac{(-1)^{i+1}kI}{4}}{1 - z} - (\rho + \sigma)^2\right]X^i(z)=0.
		\label{X-z-eq-2}
	\end{eqnarray}
	
	Comparing with the standard hypergeometric equation \cite{Duverney:2024qzj,Michel2007}
	\begin{eqnarray}
		z(1-z)\frac{d^2y}{dz^2}-\left[(a+b+1)z-c\right]\frac{dy}{dz}-aby=0,
		\label{Standard-hg}
	\end{eqnarray}
	we can find if we take the following relations
	\begin{eqnarray}
		&&
		\rho^2 - \frac{\rho}{2} + \frac{(-1)^{i+1}kI}{4}=0
		\Rightarrow
		\rho=\frac{1-\sqrt{1-4(-1)^{i+1}kI}}{4},
		\\
		&&
		\sigma^2 - \frac{\sigma}{2} - \frac{(-1)^{i+1}kI}{4}=0
		\Rightarrow
		\sigma=\frac{1+\sqrt{1+4(-1)^{i+1}kI}}{4},
	\end{eqnarray}
	\eqref{X-z-eq-2} can be reduced to standard hypergeometric equation with the coefficients
	\begin{eqnarray}
		a=b=\frac{1}{2}+\frac{\sqrt{1+4(-1)^{i+1}kI}}{4}-\frac{\sqrt{1-4(-1)^{i+1}kI}}{4}, \quad
		c=1-\frac{\sqrt{1-4(-1)^{i+1}kI}}{2}.
	\end{eqnarray}
	
	Since the standard hypergeometric equation \eqref{Standard-hg} has two linearly independent hypergeometric solutions
	\begin{eqnarray}
		y(z)=c_1 \ {}_2F\left(a, b; c; x\right) + c_2 \  z^{1-c}{}_2F\left(a-c+1, b-c+1; 2-c; x\right),
	\end{eqnarray}
	\eqref{eq-S-t} can also possess two linearly independent hypergeometric solutions, which are written in \eqref{2sol-X-t-1}-\eqref{alpha-bar}.
	
	\section{Details of the calculation for the asymptotic behavior in strongly asymptotically free case}\label{Cal-SAF}
	
	For convenience we first set $z=\frac{1-It}{2}$. Then, the two linearly independent solutions in \eqref{two-Hg-t} can be written as
	\begin{eqnarray}
		X^{i}_{\pm}(z)=(-z)^{\frac{1\mp\alpha^i}{4}} (1-z)^{\frac{1+\bar{\alpha}^i}{4}} \ F^i_{\pm}(z), \label{two-Hg-z} 
	\end{eqnarray}
	with
	\begin{eqnarray}
		F^i_{\pm}(z)=
		{}_2F
		\left(
		\frac{\bar{\alpha}^i}{4}\mp\frac{\alpha^i}{4}+\frac{1}{2}, \ \frac{\bar{\alpha}^i}{4}\mp\frac{\alpha^i}{4}+\frac{1}{2}; \ 
		1\mp\frac{\alpha^i}{2}; \ 
		z 
		\right).		
		\label{Hg-z}
	\end{eqnarray}
	
	When $t\to\infty$, we have $z\to-\infty$, then \eqref{two-Hg-z} can be written as
	\begin{eqnarray}
		\lim_{z\to-\infty}X^{i}_{\pm}(z)\propto(-z)^{\frac{1}{2}+\frac{\delta^i_{\mp}}{4}} \ \lim_{z\to-\infty}F^i_{\pm}(z), \quad \delta^i_{\mp}=\bar{\alpha}^i\mp\alpha^i. \label{lim-two-Hg-z} 
	\end{eqnarray}
	
	Comparing \eqref{rela-Hg-1} with \eqref{Hg-z}, we have 
	\begin{eqnarray}
		a^i_{\pm}=b^i_{\pm}=\frac{1}{2}+\frac{\delta^i_{\mp}}{4}, \quad c^i_{\pm}=1\mp\frac{\alpha^i}{2},
		\label{para-X1-Ax}
	\end{eqnarray}
	i.e. the formulas in \eqref{para-X1}.
	
	Substituting \eqref{rela-Hg-1}-\eqref{rela-Hg-2} into \eqref{lim-two-Hg-z}, we have
	\begin{eqnarray}
		\lim_{z\to-\infty}F^i_{\pm}(z)
		=
		\frac{\Gamma(c^i_{\pm})\Gamma(b^i_{\pm}-a^i_{\pm})}{\Gamma(b^i_{\pm})\Gamma(c^i_{\pm}-a^i_{\pm})}
		(-z)^{-a^i_{\pm}}+
		\frac{\Gamma(c^i_{\pm})\Gamma(a^i_{\pm}-b^i_{\pm})}{\Gamma(a^i_{\pm})\Gamma(c^i_{\pm}-b^i_{\pm})}
		(-z)^{-b^i_{\pm}}.
		\label{lim-F-z-1}
	\end{eqnarray}
	However, \eqref{para-X1-Ax} gives us $b^i_{\pm}-a^i_{\pm}=0$, which means the above formula arises singularities.  
	
	To regularize this divergence, we employ an infinitesimal parameter $\epsilon$ to study the limiting behavior of the Gamma function near vanishing argument.
	
	We assume 
	\begin{eqnarray}
		\epsilon=b^i_{\pm}-a^i_{\pm},
		\label{b-a}
	\end{eqnarray}
	Then by using \eqref{para-X1-Ax} and \eqref{b-a}, \eqref{lim-F-z-1} can be rewritten as
	\begin{eqnarray}
		&&
		\lim_{z\to-\infty}F^i_{\pm}(z)
		=
		\lim_{\substack{\epsilon\to0 \\ z\to-\infty }}
		(-z)^{-a^i_{\pm}}
		\left[
		\frac{\Gamma(c^i_{\pm})\Gamma(\epsilon)}{\Gamma(a^i_{\pm}+\epsilon)\Gamma(c^i_{\pm}-a^i_{\pm})}
		+
		\frac{\Gamma(c^i_{\pm})\Gamma(-\epsilon)}{\Gamma(a^i_{\pm})\Gamma(c^i_{\pm}-a^i_{\pm}-\epsilon)}
		(-z)^{-\epsilon}
		\right] \nonumber 
		\\ 
		&&
		\label{lim-F-z-2}
	\end{eqnarray}
	
	The two terms involving $\Gamma(a^i_{\pm}+\epsilon)$ and $\Gamma(c^i_{\pm}-a^i_{\pm}-\epsilon)$ can be expanded as:
	\begin{eqnarray}
		&&
		\frac{1}{\Gamma(a^i_{\pm}+\epsilon)}=\frac{1}{\Gamma(a^i_{\pm})}\left[1-\epsilon\psi(a^i_{\pm})+O(\epsilon^2)\right], 
		\label{Gamma-a-1}
		\\
		&&
		\frac{1}{\Gamma(c^i_{\pm}-a^i_{\pm}-\epsilon)}=\frac{1}{\Gamma(c^i_{\pm}-a^i_{\pm})}\left[1+\epsilon\psi(c^i_{\pm}-a^i_{\pm})+O(\epsilon^2)\right],
		\label{Gamma-c-a-1}
	\end{eqnarray}
	where $\psi(x)=\frac{\dot{\Gamma}(x)}{\Gamma(x)}$, and $\dot{(\cdot)}$ means $\frac{d}{dx}$.
	
	By using the identity of Gamma function \cite{Duverney:2024qzj}
	\begin{eqnarray}
		\Gamma(x)=\frac{\Gamma(x+1)}{x},
		\label{Gamma-x-x1}
	\end{eqnarray}
	we can expand the function $\Gamma(\epsilon)$ as
	\begin{eqnarray}
		\Gamma(\epsilon)
		=\frac{1}{\epsilon}+\dot{\Gamma}(1)+\frac{\ddot{\Gamma}(1)}{2}\epsilon+O(\epsilon^2),
		\label{z-infty-Gamma}
	\end{eqnarray}
	
	Then considering 
	\begin{eqnarray}
		(-z)^{-\epsilon}=e^{\ln\left[(-z)^{-\epsilon}\right]}=e^{-\epsilon\ln(-z)},
	\end{eqnarray}
	we have
	\begin{eqnarray}
		\lim_{\epsilon\to0}(-z)^{-\epsilon}=1-\epsilon\ln(-z)+O(\epsilon^2).
		\label{lim-ep-z}
	\end{eqnarray}
	
	Substituting \eqref{Gamma-a-1}, \eqref{Gamma-c-a-1}, \eqref{z-infty-Gamma} and \eqref{lim-ep-z} into \eqref{lim-F-z-2}, we have\footnote{Here we have dropped all of the terms for $O(\epsilon)$.}
	\begin{eqnarray}
		\lim_{z\to-\infty}F^i_{\pm}(z)
		=\frac{\Gamma(c^i_{\pm})}{\Gamma(a^i_{\pm})\Gamma(c^i_{\pm}-a^i_{\pm})}
		(-z)^{-a^i_{\pm}}\left[\ln(-z)+\mathcal{D}^i_{\pm}\right].
		\label{linear-lnz}
	\end{eqnarray}
	where $\mathcal{D}^i_{\pm}=2\psi(1)-\psi(a^i_{\pm})-\psi(c^i_{\pm}-a^i_{\pm})$. 
	
	Then substituting \eqref{linear-lnz} and \eqref{para-X1-Ax} into \eqref{lim-two-Hg-z}, we have 
	\begin{eqnarray}
		\lim_{z\to-\infty}X^i_{\pm}(z)=\frac{\Gamma(c^i_{\pm})}{\Gamma(a^i_{\pm})\Gamma(c^i_{\pm}-a^i_{\pm})}\left[\ln(-z)+\mathcal{D}^i_{\pm}\right],
		\label{lim-z-Hgz}
	\end{eqnarray}
	
	Then by using $z=\frac{1-it}{2}$, the asymptotic behavior of $X^i_{\pm}(t)$ can be written as
	\begin{eqnarray}
		\lim_{t\to\infty}X^i_{\pm}(t)=
		A_{\pm}^i+B_{\pm}^i\ln t, 
		\label{lim-t-X}
	\end{eqnarray}
	where
	\begin{eqnarray}
		A_{\pm}^i=B_{\pm}^i\left(\mathcal{D}^i_{\pm}+\frac{\pi}{2}I-\ln2\right), \quad 
		B_{\pm}^i=\frac{\Gamma(c^i_{\pm})}{\Gamma(a^i_{\pm})\Gamma(c^i_{\pm}-a^i_{\pm})}.
	\end{eqnarray}
	
	\section{Calculation of the finite displacement in the displacement memory effect for the Scarf profile}\label{cal-DM-Scarf}
	
	Now, let us discuss why $B^i_+=0$ does not imply $A^i_+=0$ in the Scarf profile.
	
	We again introduce an infinitesimal parameter $\delta$ to place the argument of the Gamma function $\Gamma(z)$ near a negative integer 
	\begin{eqnarray}
		z=-n+\delta.
	\end{eqnarray}
	Then the recurrence relation of Gamma function \cite{DLMF,Duverney:2024qzj},
	\begin{eqnarray}
		\Gamma(z+1)=z\Gamma(z),
	\end{eqnarray}
	gives us 
	\begin{eqnarray}
		\Gamma(1+\delta)=\delta(-1+\delta)(-2+\delta)\cdot\cdot\cdot(-n+\delta)\Gamma(-n+\delta).
	\end{eqnarray}
	Since $\Gamma(1)=1$, then we have
	\begin{eqnarray}
		\frac{1}{\Gamma(-n+\delta)}\sim(-1)^nn!\delta.
		\label{Gamma-nn2}
	\end{eqnarray}
	Substituting \eqref{Gamma-nn2} into the formula of $B^i_+$ in \eqref{DM-X1-AB} we have
	\begin{eqnarray}
		B^i_+\sim\frac{\Gamma(a^i_+-n)}{\Gamma(a^i_+)} (-1)^nn!\delta.
		\label{DM-Bn}
	\end{eqnarray}
	
	Then by employing the recurrence relation of $\psi(x)$ \cite{DLMF},
	\begin{eqnarray}
		\psi(x+1)=\psi(x)+\frac{1}{x},
	\end{eqnarray}
	we have 
	\begin{eqnarray}
		\psi(1+\delta)=\frac{1}{\delta}+\sum_{m=1}^{n}\frac{1}{-n+\delta}+\psi(-n+\delta),
	\end{eqnarray}
	then
	\begin{eqnarray}
		\psi(c^i_+-a^i_+)=\psi(-n+\delta)\sim \psi(1)-\frac{1}{\delta}+H_n, \quad H_n=\sum_{m=1}^{n}\frac{1}{m}.
		\label{DM-psi-n}
	\end{eqnarray}
	Substituting \eqref{DM-psi-n} into the formula of $\mathcal{D}^i_+$ in \eqref{D-psi-func}, we have
	\begin{eqnarray}
		\mathcal{D}^i_+\sim\frac{1}{\delta}+\psi(1)-\psi(a^i_+)-H_n.
		\label{DM-Dn}
	\end{eqnarray}
	
	Substituting \eqref{DM-Bn} and \eqref{DM-Dn} into the formula of $A^i_+$ in \eqref{DM-X1-AB}, we have
	\begin{eqnarray}
		A^i_+\sim \frac{\Gamma(a^i_+-n)}{\Gamma(a^i_+)} (-1)^nn!,
	\end{eqnarray}
	which means the value of $A^i_+$ is finite.
	
	\section{Details of the calculation for the asymptotic behavior in weakly asymptotically free case}\label{Cal-WAF}
	
	When $U\to\infty$, we have $z\to-\infty$, then \eqref{U3-Xpm-z} can be written as
	\begin{eqnarray}
		\lim_{z\to-\infty}X^i_{\pm}(z)
		=
		(-z)^{\tilde{b}^i_{\pm}}\lim_{z\to-\infty}F^i_{\pm}(z), \quad \tilde{b}^i_{\pm}=1+\frac{\beta^i\mp\bar{\beta}^i}{2}.
		\label{lim-U3-Xpm-z}
	\end{eqnarray}
	
	Then by employing \eqref{rela-Hg-1} again, $F^i_{\pm}(z)$ becomes
	\begin{eqnarray}
		&&
		F^i_{\pm}(z)
		=
		\mathcal{A}^i_{\pm} \cdot 
		(-z)^{-\tilde{a}^i_{\pm}}
		F^i_1(1/z)
		+
		\mathcal{B}^i_{\pm} \cdot 
		(-z)^{-\tilde{b}^i_{\pm}}
		F^i_2(1/z),
		\label{U3-lim-F-z-1}
	\end{eqnarray}
	where
	\begin{eqnarray}
		&&
		F^i_1(1/z)={}_2F\left(\tilde{a}^i_{\pm},1-\tilde{c}^i_{\pm}+\tilde{a}^i_{\pm}; \ 1-(\tilde{b}^i_{\pm}-\tilde{a}^i_{\pm}); \ \frac{1}{z}\right), 
		\\[8pt]
		&&
		F^i_2(1/z)={}_2F\left(\tilde{b}^i_{\pm},1-\tilde{c}^i_{\pm}+\tilde{b}^i_{\pm}; \ 1+(\tilde{b}^i_{\pm}-\tilde{a}^i_{\pm}); \ \frac{1}{z}\right),
	\end{eqnarray}
	and
	\begin{eqnarray}
		\mathcal{A}^i_{\pm}=\frac{\Gamma(\tilde{c}^i_{\pm})}{\Gamma(\tilde{b}^i_{\pm})\Gamma(\tilde{c}^i_{\pm}-\tilde{a}^i_{\pm})}, \quad
		\mathcal{B}^i_{\pm}=\frac{\Gamma(\tilde{c}^i_{\pm})\Gamma(\tilde{a}^i_{\pm}-\tilde{b}^i_{\pm})}{\Gamma(\tilde{a}^i_{\pm})\Gamma(\tilde{c}^i_{\pm}-\tilde{b}^i_{\pm})},
		\label{U3-A0-B0}
	\end{eqnarray}
	with
	\begin{eqnarray}
		\tilde{a}^i_{\pm}=\frac{\beta^i\mp\bar{\beta}^i}{2}, \quad
		\tilde{c}^i_{\pm}=1\mp\bar{\beta}^i.
	\end{eqnarray}
	
	Note that both terms on the right‑hand side in \eqref{U3-lim-F-z-1} possess singularities:
	\begin{itemize}
		\item
		In the first term, the parameter $1-(\tilde{b}^i_{\pm}-\tilde{a}^i_{\pm})=0$ in $F^i_1(1/z)$ leads to the fact that the identity \eqref{rela-Hg-2} fails for $F^i_1(1/z)$ but holds only for $F^i_2(1/z)$.
		
		\item 
		In the second term, the Gamma function $\Gamma(\tilde{a}^i_{\pm}-\tilde{b}^i_{\pm})=\Gamma(-1)$ in $\mathcal{B}^i_{\pm}$ is singular.
		
	\end{itemize}
	We need to treat these two singularities specially.
	
	The hypergeometric function ${}_2F(a,b;c;x)$ can be expanded in power of $x$ \cite{Michel2007}
	\begin{eqnarray}
		&&
		{}_2F(a,b;c;x)=\sum_{n=0}^{\infty}\frac{(a)_n(b)_n}{(c)_nn!}x^n=\sum_{n=0}^{\infty}t_nx^n, 
		\label{power-z-2F-1}
		\\[8pt]
		&&
		t_0=1, \quad t_{n+1}=\frac{(a+n)(b+n)}{(c+n)(n+1)}t_n, \quad n\geq0.
		\label{power-z-2F-2}
	\end{eqnarray}
	This will be the key to our treatment of the singularity in the first term.
	
	We also introduce an infinitesimal quantity $\epsilon$,
	\begin{eqnarray}
		\epsilon=1-(\tilde{b}_{\pm}-\tilde{a}_{\pm}).
		\label{U3-epsilon}
	\end{eqnarray}
	Then by using \eqref{power-z-2F-1}-\eqref{power-z-2F-2}, $F^i_1(1/z)$ can be expanded in power of $\frac{1}{z}$
	\begin{eqnarray}
		F^i_1(1/z)=1-\frac{\tilde{a}^i_{\pm}(\tilde{c}^i_{\pm}-\tilde{a}^i_{\pm}-1)}{\epsilon}\frac{1}{z}+O(z^{-2}).
		\label{power-z-2F1}
	\end{eqnarray}
	Then substituting \eqref{U3-epsilon}-\eqref{power-z-2F1} into \eqref{U3-lim-F-z-1} and after a long and complicated calculations, we obtain
	\begin{eqnarray}
		&&
		F^i_{\pm}(z)
		=
		\lim_{\epsilon\to0}
		\left[
		(-z)^{-\tilde{b}^i_{\pm}}\mathcal{A}^i_{\pm}
		\left(
		(-z)^{1-\epsilon}
		+
		p^i \frac{(-z)^{-\epsilon}}{\epsilon}
		+p^i \ \Gamma(\epsilon-1) 
		\right)
		\right]
		+O(z^{-2}),
		\label{U3-lim-F-z-2}
	\end{eqnarray}
	where
	\begin{eqnarray}
		p^i=a^i_{\pm}(c^i_{\pm}-a^i_{\pm}-1)=\frac{(-1)^{i}kI}{2}.
	\end{eqnarray}
	
	Now let us move to the second singularity.
	
	By using \eqref{Gamma-x-x1} and \eqref{z-infty-Gamma}, the Gamma function $\Gamma(\epsilon-1)$ can be expanded in power of $\epsilon$
	\begin{eqnarray}
		\Gamma(\epsilon-1)=-\frac{1}{\epsilon}-m+O(\epsilon), \quad m=\psi(1)+1.
		\label{exp-Gamma-epsilon}
	\end{eqnarray} 
	Then substituting \eqref{exp-Gamma-epsilon} into \eqref{U3-lim-F-z-2}, we have
	\begin{eqnarray}
		F^i_{\pm}(z)
		=
		\lim_{\epsilon\to0}
		\left[
		(-z)^{-\tilde{b}^i_{\pm}}\mathcal{A}^i_{\pm}\left[ (-z)-p^i\ln(-z)\right]+M^i_{\pm}+O(z^{-2})+O(\epsilon)
		\right],
	\end{eqnarray}
	where $M^i_{\pm}$ is a constant, which contains a collection of constant terms whose precise form is irrelevant, and we therefore omit its expression.
	
	From the above, the asymptotic expression in \eqref{lim-U3-Xpm-z} finally becomes
	\begin{eqnarray}
		\lim_{z\to-\infty}X^i_{\pm}(z)
		=
		\mathcal{A}^i_{\pm}\left[(-z)-p^i\ln(-z)\right]+M^i_{\pm}.
		\label{lim-U3-Xpm-z-final-Ax}
	\end{eqnarray}
	
\end{document}